\documentclass[pdflatex,sn-basic]{sn-jnl}

\usepackage{graphicx}%
\usepackage{multirow}%
\usepackage{amsmath,amssymb,amsfonts}%
\usepackage{amsthm}%
\usepackage{mathrsfs}%
\usepackage[title]{appendix}%
\usepackage{xcolor}%
\usepackage{textcomp}%
\usepackage{manyfoot}%
\usepackage{booktabs}%
\usepackage{algorithm}%
\usepackage{algorithmicx}%
\usepackage{algpseudocode}%
\usepackage{listings}%
\usepackage{bm}
\usepackage{enumitem}
\usepackage{array}

\begin{document}

\title{Modeling the impact of thermal stresses induced by wellbore cooldown on the breakdown pressure and geometry of a hydraulic fracture}


\author[1]{\fnm{Guanyi} \sur{Lu}}\email{gul51@pitt.edu}

\author[2]{\fnm{Mark} \sur{Kelley}}\email{kelleym@battelle.org}

\author[2]{\fnm{Samin} \sur{Raziperchikolaee}}\email{raziperchikolae@battelle.org}

\author*[1,3]{\fnm{Andrew} \sur{Bunger}}\email{bunger@pitt.edu}

\affil[1]{\orgdiv{Department of Civil and Environmental Engineering}, \orgname{University of Pittsburgh}, \orgaddress{\city{Pittsburgh}, \state{PA}, \country{USA}}}

\affil[2]{\orgname{Battelle Memorial Institute}, \orgaddress{\city{Columbus}, \state{OH}, \country{USA}}}

\affil[3]{\orgdiv{Department of Chemical and Petroleum Engineering}, \orgname{University of Pittsburgh}, \orgaddress{\city{Pittsburgh}, \state{PA}, \country{USA}}}


\abstract{Wellbore cooldown is often employed before well stimulation and/or hydraulic fracture stress testing in Enhanced Geothermal Systems and high temperature petroleum reservoirs to prevent equipment from being overheated due to high temperatures. The thermo-elastic stress resulting from heat conduction during the cooling activity can have important influence on the behavior of the hydraulic fractures. A coupled numerical model has been developed to study the thermo-mechanical effect associated with pre-injection wellbore cooldown on the wellbore pressure and geometry of the hydraulic fracture (either longitudinal or transverse to the wellbore axis). The main novelty of this numerical study is the consideration of significant near-wellbore thermal stresses in the coupled non-linear problem of hydraulic fracturing initiation and propagation, which enables investigation of the thermo-mechanical effect under different fracture propagation regimes. Simulation results show earlier fracture initiation and lower breakdown pressure caused by cooling circulation. Extensive wellbore cooling also significantly alters the evolution of wellbore pressure, as evidenced by the differences observed under various cooling conditions. Most importantly, cooling promotes the transverse initiation of hydraulic fractures in situations where the initiation would have been longitudinal (i.e. in the same plane as the well) in the absence of cooling. The cases most susceptible to the complete change of fracture initiation geometry are those where the well is drilled parallel to the least compressive stress, typically horizontal wells drilled parallel to the minimum horizontal stress but also applicable to vertical wells in cases where the vertical stress is the lower in magnitude than either horizontal principal stress. These results combine to indicate a profound potential for cooling to impact hydraulic fracture initiation and early growth, and therefore needs to be considered in the planning and interpretation of stress testing and reservoir stimulation when cooling operations are necessary. }




\maketitle

Highlights
\begin{itemize}
\item A hydraulic fracturing model coupled with heat conduction induced by cooling circulation is constructed to study the impact of wellbore cooldown.
\item Wellbore cooldown leads to an earlier hydraulic fracture initiation with a lower breakdown pressure, as well as reduces the wellbore pressure during fracture growth.
\item Cooling promotes the transverse initiation of the hydraulic fracture from wells drilled along the minimum in-situ stress direction.
\end{itemize}

\section{Introduction}\label{sec_intro}

Hydraulic fracturing stimulation techniques have been utilized in Enhanced Geothermal Systems (EGS) to facilitate the exploitation of the geothermal resources \citep{MuTe81,ZiMo10,McHo14,KeWo15,MoMc20}. High permeability fracture pathways are created by hydraulically stimulating the hot rock formation at depth. In a typical EGS geothermal reservoir, the temperature can easily exceed 200 $^\circ$C \citep{GhTa07,BrDu12,KeWo15,MoMc20}. The excessive temperature poses serious challenges to the drilling operation and the usage of electronic data acquisition devices \citep{SiJo11,BrDu12,XiHu22}. Therefore, extensive wellbore cooldown by circulating cold fluid over a period of time in the borehole before lowering down the tools is often necessary for preventing the devices from being overheated \citep{XiHu22}. When the cold fluid is being circulated in the borehole, it rapidly cools the near-wellbore rock volume due to the large temperature difference between the borehole (mainly controlled by the cold fluid) and the hot rock formation. This cooling process, sometimes lasting for days \citep{BrDu12}, would inevitably induce thermal stresses in areas adjacent to the wellbore \citep{StVo82,PeGo84,WaPa94,LiCu98,GhZh04,Chen16}. As a result, the stress distributions (both the tangential and the axial stresses) in the region adjacent to the wellbore are substantially altered.

For hydraulic fracturing treatments, the near-wellbore stress field plays a critical role in the determination of the breakdown pressure \citep{HuWi57,HaFa67,DeCa97,BuLu15,LaDe16,LuGo17}, and it can potentially affect the fracture geometry during the initial stage of pressurization \citep{LeAb13,BeDj21}. In the particular case of mini-frac (mini-HF) testing, where small scale hydraulic fractures (HFs) are created for in-situ stress measurements \citep{CoVa84,DeWa89,HaCo03}, the near-field thermo-elastic stresses induced by wellbore cooling can have a dominating effect on the pressure evolution and/or the orientation of the induced fracture. Failure to account for the thermally induced stresses would certainly lead to erroneous interpretation of the test results. 

In recent decades, a growing body of research has investigated the impact of heat conduction on the hydraulic fracturing process \citep{StVo82,TrRo10,LeGh11,ToGu17,ZhJi21}. The thermally induced stresses (tension in the case of cold fracturing fluid) are shown to result in notable changes in the fluid pressure required to break the rock and to propagate the HFs. These pioneering works have undoubtedly provided valuable insights into the thermo-mechanical effects by the cold injection fluid on the growth of a fluid driven hydraulic fracture. However, the impact of substantial wellbore cooldown prior to the onset of hydraulic fracturing has not been thoroughly examined. More importantly, the orientation of the fracture is generally predefined in the past efforts to model the HF's evolution. It remains unclear how such cooling process at the borehole affects the favorable geometry of the fracture when being initiated during pressurization.

In this work, a fully coupled numerical solver for the initiation and early-time propagation of a planar HF by \citet{LuGo17,LuGo18} is adopted. The fracture is allowed to initiate and propagate either in the longitudinal direction parallel to the wellbore axis, or along the transverse direction perpendicular to the wellbore. The additional cooling induced stresses are taken into account via the elasticity equation to study the roles of different cooling parameters, such as temperature difference and duration of cooldown. The problem formulation is given in Section \ref{sec_prob}. Section \ref{sec_methods} presents details of the numerical methods. Simulation results and their implications for hydraulic fracturing in EGS are discussed in Section \ref{sec_results}. Finally, conclusions from this study are drawn in Section \ref{sec_conclusions}. Note that in all cases, the conditions are taken to be typical of EGS reservoirs, but the principles also apply to high temperature petroleum reservoirs where cooling operations are also necessary prior to stimulation and/or mini-HF stress testing.

\section{Problem formulation}\label{sec_prob}

\subsection{Model description}\label{subsec_descrip}

The goal of this work is to examine the effect of wellbore cooling on the subsequent hydraulic fracturing process. Past experimental and theoretical studies have shown that tension induced by cooling can lead to the occurrence of purely thermally induced/driven cracks \citep{DoGh05,TaGh11,TaGh12b,TaGh14,GrLe18,ChAl18,ChZh22,LuCh22}, or thermally aided failure/cracks \citep{BrZo99,BeCo03,TaGh10,TaGh12a} in rocks. This scenario occurs under conditions of relatively low compressive stresses, significant temperature variation, and/or the presence of other contributors such as mud pressure. In the case of well cooldown, the fluid circulation does not necessarily lead to thermal cracking in typical EGS formations under relatively high in-situ stresses at depth \citep{ToGu17}. In this study, we focus on the situations in which thermally induced stresses alone are insufficient to create fractures, but they act together with the fluid pressure, which is considered the driving force for HF initiation and propagation, to facilitate the process. 

\begin{figure}[ht]
\begin{centering}
\includegraphics[width=\columnwidth]{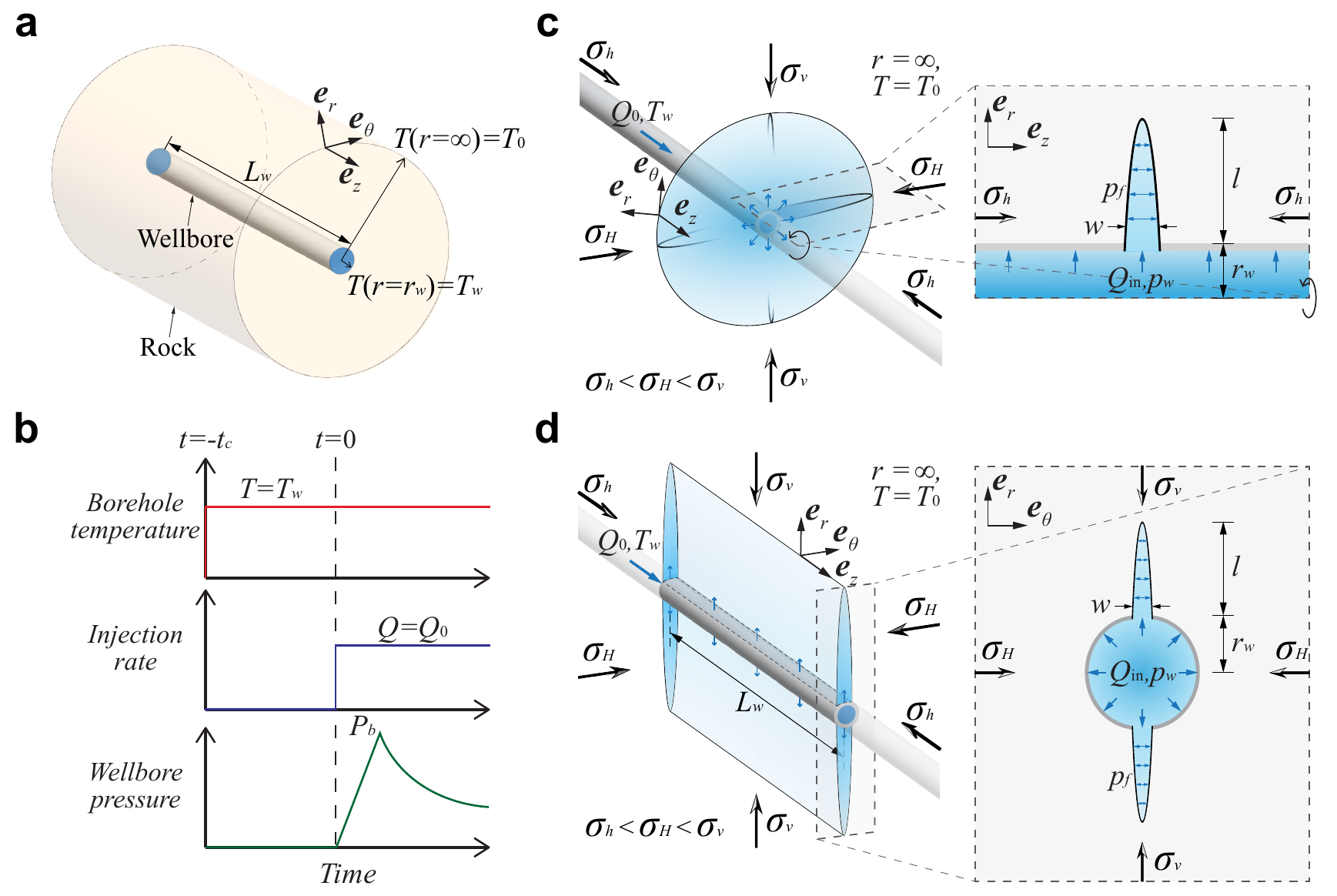}
\par\end{centering}
\caption{a, One-dimensional (1-D) axisymmetric heat diffusion driven by a constant temperature at the wellbore interval with length $L_w$. b, Illustration of the constant temperature (top) and fluid injection (middle) boundary conditions at the wellbore wall and the theoretical evolution of wellbore pressure (bottom) in response to fluid injection.  c, 3-D sketch of a transverse HF perpendicular to the horizontal wellbore (left) and planar view ($\bm{e}_z\bm{e}_r$-plane) of the axisymmetric HF from the wellbore with radius $r_w$. d, A longitudinal HF growing perpendicular to the intermediate in-situ stress ($\sigma_H$) direction (left) and the cross sectional view of the plane-strain bi-wing fracture on $\bm{e}_\theta\bm{e}_r$-plane (right).\label{fig:setup}}
\end{figure}

The problem is thus divided into two sub-problems based on time $t$: (1) When $-t_c \leq t \leq 0$, cold fluid is circulated along the wellbore interval (\hyperref[fig:setup]{Fig.~\ref*{fig:setup}a}) to cool down the borehole and near-field rock volume. Specifically, cooling starts at $t=-t_c$ and ends at $t=0$ with a constant temperature difference between the cold fluid at wellbore wall, $T_w$, and rock far away from the wellbore, $T_0$, given as $T_1=T_w-T_0$); (2) Fracturing fluid is injected into the borehole (starts immediately after the cooling process at $t = 0$) to initiate and propagate a HF (either a transverse or a longitudinal fracture as shown in \hyperref[fig:setup]{Fig.~\ref*{fig:setup}c and d}). The temperatures at the wellbore and rock formation in the far-field remain unchanged as $T_w$ and $T_0$, respectively. The constant temperature and fluid injection rate conditions are illustrated in \hyperref[fig:setup]{Fig.~\ref*{fig:setup}b}, along with an idealized plot of the wellbore pressure ($P_w$) evolution.

It is worth noting that in the cooldown period before hydraulic fracturing as well as during the fracturing process, the heat conduction that occurs between the wellbore and the reservoir both contribute to the rise of thermo-elastic stresses in near-wellbore regions. Since the fluid volume flowing into the fracture is relatively small, and the fluid would be instantly heated up when it enters the fracture, we neglect the heat transfer between the newly created fracture surface and the reservoir. This assumption is justified for early-time fracture initiation and propagation. For HFs with much larger magnitudes compared to the wellbore size, the overall behavior will certainly be governed by the temperature of fluid inside the fracture, instead of its value at wellbore. In these situations, an approach that considers transient solutions of the thermo-poro-elastic responses of fractures, such as \citet{GhZh06} and \citet{ToGu17}, is more suitable. 

In a fully coupled thermo-mechanical model, the exchange of heat causes the solid to deform, and solid deformation also generates heat. According to \citet{Chen16}, the heat generated by solid deformation is negligible when ${K\beta_d^2}/{m_d} \ll 1$, where $K$ is the drained bulk modulus, $\beta_d$ is the drained coefficient of volumetric thermal expansion for porous medium frame, and $m_d$ gives the value of a drained specific heat at constant strain under a reference state ($c_d$) divided by the reference temperature (293 K), $m_d={c_d}/{293}$. This correlation holds true for most granitic rocks, which are commonly encountered in EGS \citep{McHo14,KeWo15,MoMc20}. For instance, $\beta_d=2.4 \times 10^{-5}\;\text{K}^{-1}, m_d=7.85 \times 10^3\;\text{J}/\text{m}^3, K=2.5 \times 10^{10}\;\text{Pa}$ for Westerly granite \citep{Chen16}, resulting in ${K\beta_d^2}/{m_d}=0.0018 \ll 1$. Clearly, the heat induced by rock deformation is much smaller compared to the heat being transported into the reservoir from the wellbore wall during cooling and hydraulic fracturing. Hence, we neglect the heat caused by elastic deformation in our model. To summarize, the thermo-mechanical part is characterized by the axisymmetric heat conduction between the borehole and the reservoir with a constant temprature difference, $T_1$, and its associated time varying thermally induced stress fields. 

Heat diffusion can possibly induce pore pressure change in a saturated rock. However, after hours to days of cooling circulation, the rock reaches a drained condition and any thermally induced pore pressure would be dissipated \citep{Chen16}. For the second sub-problem of hydraulic fracturing, the low permeability of granitic rocks in typical EGS projects justifies the assumption of zero fluid leak-off and its associated pore pressure diffusion. Therefore, pore pressure is considered negligible in both sub-problems. Fluid lag at the fracture tip is assumed to be negligible, and the rock is considered as an isotropic material in our model. Additionally, laboratory experiments revealed that for a granitic rock below 300 $^\circ$C, temperature has limited impact on its mechanical properties \citep{LiMa20}. Considering the temperature range for the majority of current EGS projects (mostly below 300 $^\circ$C), the rock's mechanical properties, including the elastic modulus $E$, Poisson's ratio $\nu$, drained bulk modulus $K$, tensile strength $\sigma_T$, fracture toughness $K_\text{IC}$, and thermo-mechanical properties (detailed in Section \ref{subsec_goveq}) are all set to be constant. 

Evidences from both laboratory experiments \citep{WeDe94,Weij95,AbHe96,KeWh13,LuUw15,LeDe17,LuGo20,LiLe20,LuMo22,LiLe22a,WaTa23} and field applications \citep{WaHe06,Dane11} suggest that a HF may be initiated either along the perpendicular direction with respect to the wellbore axis (hereafter referred to as the transverse hydraulic fracture), or on the plane parallel to the wellbore axis (longitudinal orientation). Take a horizontal well drilled along the direction of minimum in-situ stress, $\sigma_h$, as an example, the transverse fracture corresponds to an axisymmetric fracture of length $l$ emanating from a horizontal well with radius $r_w$ under minimum in-situ stress, $\sigma_h$, acting orthogonal to the fracture plane (\hyperref[fig:setup]{Fig.~\ref*{fig:setup}c}). On the other hand, the longitudinal fracture (\hyperref[fig:setup]{Fig.~\ref*{fig:setup}d}) is defined as a bi-wing fracture under plane strain conditions under the bi-axial far-field compressive stresses (intermediate in-situ stress, $\sigma_H$, and maximum in-situ stress, $\sigma_v$). Due to near-wellbore stress concentration and solid-fluid coupled hydraulic fracturing mechanics (propagation regimes and their impact on the pressure evolution as discussed in previous works such as \citet{Deto16,LaDe16,LuGo17}), the HF is not necessarily initiated perpendicular to the minimum in-situ stress ($\sigma_h$) direction. In this numerical study, we focus on these two most probable scenarios (either a transverse or a longitudinal hydraulic fracture initiation from a well drilled parallel to the direction of minimum in-situ stress). Whether the transverse or the longitudinal direction is the favorable fracture geometry depends on the stress state, as well as fluid injection conditions \citep{Weij95,LeAb13,BeDj21}. To study how thermally induced stresses affect the fracture orientation, we adopt the method by \citet{LeAb13} - the orientation of the HF is determined by whichever geometry requires lower pressure to break the rock (i.e., breakdown pressure $P_b$). 

In order to do so, both fractures are set to start with an initial defect of the same length $l_0$. A Newtonian fluid with viscosity $\mu$ is injected at a constant rate $Q_0$ into the wellbore interval of length $L_w$ with compressibility $U$. Initially, a very small fluid pressure is imposed on the entire starter crack. The continuous injection results in the increase in both fracture width and fluid pressure within the stater crack while the crack length remains unchanged during this pressurization phase. The fracture initiation is achieved at the time when the mode I (opening) stress intensity factor, $K_\text{I}$, reaches the value of the fracture toughness, $K_\text{IC}$. Therefore, this model is capable of capturing both the time and pressure needed to initiate the HF. Then, the HF propagation, modeled by a coupled non-linear system, is solved using a length-controlled numerical scheme presented in Section \ref{sec_methods}.

\subsection{Governing equations}\label{subsec_goveq}

\subsubsection{Heat conduction}\label{subsubsec_heateqs}

For sub-problem 1, the temperature field is governed by 1-D axisymmetric heat diffusion equation \citep{Chen16}

\begin{equation}
 \frac{\partial T}{\partial t^\prime}=\kappa_T\frac{1}{r}\frac{\partial}{\partial r}(r\frac{\partial T}{\partial r})\;,\;\;\; r \geqslant r_w \label{eq:diffusion_eq}
\end{equation}
where $r$ is the radial coordinate as shown in \hyperref[fig:setup]{Fig.~\ref*{fig:setup}} ($r=0$ at the center of the borehole), and $t^\prime=t+t_c$. In Eq. \eqref{eq:diffusion_eq}, $\kappa_T={k_T}/{c_d}$ is introduced as heat diffusivity coefficient, and $k_T$ and $c_d$ are the effective thermal conductivity and the drained specific heat, respectively. The temperature distribution in the surrounding area of the borehole is then determined by the following boundary and initial conditions 

\begin{equation}
\begin{aligned}
T=T_w\;\;& \mathrm{at}\;\;r=r_w\;\;\mathrm{for}\;\;t^\prime\geq0\\
T=T_0\;\;& \mathrm{at}\;\;r=\infty\\
\end{aligned}\label{eq:diff_ini_cond}
\end{equation}

\subsubsection{Linear Elasticity}\label{subsubsec_elasticity}

For the solid-fluid coupled hydraulic fracturing problem (sub-problem 2), two formulations of the elastic responses of the rock are given depending on the fracture geometry \citep{KeLu77,CrSt83,AbLe13,LuGo17,LuGo18}. For the axisymmetric (transverse) fracture from a horizontal well (\hyperref[fig:setup]{Fig.~\ref*{fig:setup}c}), the relation between the fracture width $w$ and the net pressure acting on the fracture is represented by a boundary integral equation

\begin{equation}
p_f(r,t)-\sigma_{zz}^{\Delta T}(r,t^\prime)-\sigma_n(r) = \int_{\Sigma}h(r,r^\prime)\frac{\partial w}{\partial r^\prime}\text{d}r^\prime\label{eq:elasticity_axi}
\end{equation}
in which $\sigma_n(r)=\sigma_h$ is the constant far-field compressive stress, and $\sigma_{zz}^{\Delta T}$ is the thermally induced axial stress (compression positive) normal to the crack face (along $\bm{e}_z$ direction). The integration on the right hand side of Eq. \eqref{eq:elasticity_axi} is carried over $\Sigma$, which represents the entire crack length $\Sigma=\{r^\prime:r^\prime\in\left(r_w,r_w+l\right)\}$. The elastic kernel $h(r,r^\prime)$ denotes the stress at $r$ caused by an axisymmetric dislocation of radius $r^\prime$ in an infinite domain with the cylindrical cavity of radius $r_w$ \citep{KeLu77}.

For a plane-strain (longitudinal) bi-wing HF propagating from the horizontal well interval with length $L_w$ (\hyperref[fig:setup]{Fig.~\ref*{fig:setup}d}), displacement discontinuity (DD) formulations are used \citep{CrSt83}

\begin{equation}
\begin{aligned}
p_f(r,t)-\sigma_{\theta\theta}^{\Delta T}(r,t^\prime)-\sigma_n(r) & =
\int_{\Sigma}\left[H_{nn}\left(r,r^\prime\right)u_n(r^\prime,t)+H_{ns}(r,r^\prime)u_s\left(r^\prime,t\right)\right]\text{d}r^\prime\\
0 & = \int_{\Sigma}\left[H_{sn}\left(r,r^\prime\right)u_n(r^\prime,t)+H_{ss}(r,r^\prime)u_s\left(r^\prime,t\right)\right]\text{d}r^\prime\\
\end{aligned}\label{eq:elasticity_ps}
\end{equation}

The involved kernels $H$ in Eq. \eqref{eq:elasticity_ps} represent the stresses (normal or tangential) at $r$ caused by a unit DD (normal or tangential) at point $r^\prime$. For a longitudinal HF from a horizontal well drilled in the direction of the minimum in-situ stress, the compressive stress normal to the plane of the HF growth, $\sigma_n$, is given by the Kirsch solution for near-wellbore stress concentration under bi-axial stresses $\sigma_v$ and $\sigma_H$ \citep{kirs98}

\[\sigma_{n}(r)=\sigma_{H}\left(1+\frac{r_w^2}{r^2}\right)+\frac{\sigma_v-\sigma_H}{2}\left(\frac{r_w^2}{r^2}-3\frac{r_w^4}{r^4}\right)\]

For the plane-strain configuration, the integral kernels, \textit{H}, do not account for the wellbore directly. Hence, both the wellbore wall and the HF are included in the integration contour ($\Sigma$) and are discretized by the DD elements \citep{ZhJe11,LuGo17}. For the axisymmetric fracture, the elastic kernels have already taken into account the presence of the wellbore with radius, $r_w$, and thus the integration is conducted along the path of the HF. More detailed mathematical expressions of the elastic kernels in Eqs. \eqref{eq:elasticity_axi} and \eqref{eq:elasticity_ps} can be found in \citet{LuGo17} for a single fracture and \citet{LuGo18} for multiple fractures growing simultaneously.

As a key component in the numerical model, the thermo-mechanical effect is taken into consideration in the elasticity equations \eqref{eq:elasticity_axi} and \eqref{eq:elasticity_ps} by introducing the thermally induced stress alteration (axial stress component $\sigma_{zz}^{\Delta T}$ and tangential stress component $\sigma_{\theta\theta}^{\Delta T}$) onto the existing stresses acting on the fracture faces. According to \citet{Nowa86}, \citet{GhNy08}, and \citet{Chen16}, the thermo-elastic stress alteration is uniform along all directions, which is given by

\begin{equation}
\sigma^{\Delta T}(r,t^\prime)=\alpha_d \Delta T(r,t^\prime)\label{eq:thermal_stress}
\end{equation}
where $\sigma^{\Delta T}=\sigma_{\theta\theta}^{\Delta T}=\sigma_{zz}^{\Delta T}$, $\Delta T(r,t^\prime)=T(r,t^\prime)-T_0$, and $\alpha_d$ is the drained thermo-elastic effective stress coefficient, which can be obtained by $\alpha_d=K\beta_d$.


\subsubsection{Fluid flow}

The fluid flux $q\left(r,t\right)$ in the fracture is governed by the Poiseuille law \citep{Batc67}. For an incompressible Newtonian fluid with a dynamic viscosity $\mu$, we have

\begin{equation}
q=-\frac{w^3}{\mu'}\frac{\partial p_f}{\partial r} \;\;\mathrm{for}\;\;r\in\left(r_w,r_w+l\right)\label{eq:poiseuille}
\end{equation}
where $\mu'=12\mu$. The continuity equation is derived based on the fluid mass balance

\begin{equation}
\frac{\partial w}{\partial t}+\frac{1}{r^{d-1}}\frac{\partial}{\partial r}\left(r^{d-1}q\right)=0\label{eq:continuity}
\end{equation}
with $d = 1$ for plane-strain and $d = 2$ for axisymmetry.

\subsubsection{Fracture initiation and propagation}

To solve for the initiation and propagation of a HF, we impose the quasi-static propagation condition at the crack tip, $K_\text{I}=K_\text{IC}$, in which $K_\text{I}$ is the stress intensity factor, and $K_\text{IC}$ represents the fracture toughness of the rock. This propagation condition can be expressed as an asymptote (``\textit{k}'' asymptote) of the fracture opening, $w$, at the tip of the propagating HF \citep{Irwi57}

\begin{equation}
w\sim\frac{K'}{E'}\hat{r}^{1/2}\;\;\mathrm{for}\;\;\hat{r}\rightarrow0\label{eq:k_asymptote}
\end{equation}
where $\hat{r}=r_w+l-r$ denotes the distance to the crack tip, $E'=E/\left(1-\nu^{2}\right)$ is the plane-strain elastic modulus, and $K'=\sqrt{32/\pi}K_{\mathrm{IC}}$.

In theory, cooling circulation in sub-problem 1 induces tension on the HF \citep{Chen16}. Here we recall that this study focuses on the scenario that thermal stresses do not generate enough tension to propagate the starter crack (i.e., $K_\text{I}<K_\text{IC}$ under thermally induced stresses). A uniform initial pressure, $p_i$, is then applied along the interior faces of the starter crack with length $l_0$ to ensure a non-negative opening of the fracture while the crack remains static, that is 

\begin{equation}
p_{f}\left(r\right)=\sigma_{h}+p_{i}\;\;\;\left(t=0,\;r_w\leq r\leq r_w+l_0,\;K_\text{I}<K_\text{IC}\right)\label{eq:hf_ini_cond}
\end{equation}
where the value of $p_i$ varies in different cases, depending on the magnitude of the normal stresses $\sigma_{n}+\sigma^{\Delta T}$.

\subsubsection{Boundary conditions}

For a growing HF, the crack-tip boundary condition imposes vanishing fluid flux and fracture width at the fracture tip

\begin{equation}
w(r,t)=0,\;\;q(r,t)=0\;\;\mathrm{at}\;\;r=r_w+l\label{eq:tip_bound_cond}
\end{equation}

The inlet boundary condition for the fluid flux entering the fracture from the wellbore is derived from the mass balance accounting for the injection rate and the system compressibility \citep{LhDe05sfc,AbLe13,LaDe16,LuGo17,LuMo22,LiLe22a}

\begin{equation}
q\left(r,t\right)=\frac{1}{2\left(\pi r\right)^{d-1}}\left(Q_{0}-U\frac{\text{d}P_{w}}{\text{d}t}\right)\;\;\mathrm{at}\;\;r=r_w\label{eq:inlet_bound_cond}
\end{equation}
with $P_w=p_f\left(r=r_w\right)$. For the longitudinal plane-strain HF, the injection rate $Q_0$ and compressibility $U$ in Eq. \eqref{eq:inlet_bound_cond} are taken as their values per unit length of the interval, $Q_0/L_w$ and $U/L_w$, respectively. Finally, the integration of continuity equation \eqref{eq:continuity} over time and space while accounting for both boundary conditions \eqref{eq:tip_bound_cond} and \eqref{eq:inlet_bound_cond} renders the global volume balance equation

\begin{equation}
2\int_{r_w}^{r_w+l}\left(\pi r\right)^{d-1}\left(w\left(r,t\right)-w\left(r,0\right)\right)\text{d}r=Q_0t-U\left(P_w\left(t\right)-P_w\left(0\right)\right)\label{eq:global_vol_bal}
\end{equation}

\section{Methods}\label{sec_methods}

\subsection{Temperature field}\label{subsec_heat_sol}

The heat conduction sub-problem defined by Eqs. \eqref{eq:diffusion_eq} and \eqref{eq:diff_ini_cond} is decoupled from the solid deformation as discussed in Section \ref{subsec_descrip}. Given the constant temperature boundary conditions, the temperature field is obtained in Laplace transform domain by \citet{Chen16}. A numerical inversion method \citep{Steh70} is then applied to evaluate the solution for temperature, $T$, in the time domain. Details of the solution and the inversion method are given in Appendix \ref{sec_Appendix_temp_sol}.

\subsection{Scaling}

\subsubsection{Dimensionless variables}\label{subsec_scaled_var}

We now turn to the hydraulic fracturing problem. Scaling analysis is adopted to reduce the dimensionality of the parametric space of the hydraulic fracturing problem. Following \citet{AbLe13} and \citet{LuGo17,LuGo18}, we introduce the following dimensionless variables

\begin{equation}
\gamma=\frac{l}{L_{*}} , \tau=\frac{t}{t_{*}}, \mathit{\Pi}_{f}=\frac{p_{f}}{p_{*}}, \mathit{\Omega}=\frac{w}{w_{*}}, \mathit{\Psi}=\frac{q}{q_{*}}.\label{eq:scaling}
\end{equation}
where $L_{*}$, $t_{*}$, $p_{*}$, $w_{*}$, and $q_{*}$ are the 5 characteristic scales for crack length, time, pressure, width, and fluid flux, which are obtained as follows

\begin{equation}
L_{*}=\left(E^\prime U\right)^{\frac{1}{d+1}}, t_{*}=\frac{K^\prime U^{\frac{2d+1}{2d+2}}}{Q_{0}{E^\prime}^{\frac{1}{2d+2}}}, p_{*}=\frac{K^\prime}{\left(E^\prime U\right)^{\frac{1}{2d+2}}},
w_{*}=\frac{K^\prime U^{\frac{1}{2d+2}}}{{E^\prime}^{\frac{2d+1}{2d+2}}}, q_{*}=\frac{Q_{0}}{\left(E^\prime U\right)^{\frac{d-1}{d+1}}}.\label{eq:scaling_UK}
\end{equation}

Additionally, we define the following dimensionless variables based on the 5 characteristic scales
\begin{equation}
\mathcal{{A}}=\frac{r_w}{L_{*}}, \xi=\frac{r}{L_{*}}, \bar{\sigma}=\frac{\sigma}{p_{*}}, \mathit{\Pi}_{w}=\frac{P_{w}}{p_{*}}.\label{eq:scaling_2}
\end{equation}

After scaling, the only characteristic parameter that appears in the scaled governing equations is the dimensionless viscosity, $\mathcal{M}$, given by

\[\mathcal{M}=\frac{\mu^\prime Q_{0}{E^\prime}^{\frac{2d+4}{d+1}}}{{K^\prime}^4U^{\frac{d-1}{d+1}}}\]

In addition to the aforementioned dimensionless variables related to the HF initiation and propagation sub-problem, we introduce two dimensionless parameters associated with the wellbore cooldown conditions: 

\begin{enumerate}

\item Characteristic cooldown time
\[\mathcal{T}_c=\frac{\kappa_T}{r_w^2}t_c\]

Although the thermal stress field evolves during both the cooling circulation and hydraulic fracturing activity, the duration of HF initiation and propagation is much shorter than the preceding fluid circulation ($t\ll t_c$). Therefore, the scaled cooling time, $\mathcal{T}_c$, plays a dominating role in the distribution of induced thermo-elastic stress.

\item Characteristic thermo-elastic stress
\[\bar{\sigma}^{T_1}=\frac{\sigma^{T_1}_\ast}{p_\ast}=\alpha_d T_1\frac{\left(E^\prime U\right)^{\frac{1}{2d+2}}}{K^\prime}\]
\end{enumerate} 

Given that the cooling induced temperature field solely affects the HF's growth through the generated thermo-elastic stress, the scaling of temperature, as well as the heat diffusion equation \eqref{eq:diffusion_eq}, are deemed unnecessary.

\subsubsection{Scaled governing equations}

Here we briefly summarize the dimensionless forms of the governing equations and the boundary/initial conditions scaled using the dimensionless groups described in Section \ref{subsec_scaled_var}.

\subsubsection*{Elasticity}

For the axisymmetric dislocation formulation, the scaled elasticity equation can be derived as

\begin{equation}
\mathit{\Pi}_{f}\left(\xi,\tau\right)-{\bar{\sigma}_{zz}}^{\Delta T}\left(\xi,\tau\right)-\bar{\sigma}_{n}\left(\xi\right) = \int_{\Sigma}\bar{h}\left(\xi,\xi^\prime\right)\frac{\partial\mathit{\Omega}}{\partial\xi}\text{d}\xi^\prime\label{eq:scaled_elasticity_axi}
\end{equation}
where $\xi^\prime={r^\prime}/{L_\ast}$, and $\bar{h}$ denotes the scaled elastic kernels. The integral equation for the plane-strain HF \eqref{eq:elasticity_ps} with the DD kernels can be similarly formulated in the scaled form.

\subsubsection*{Poiseuille law}

The scaled Poiseuille law is given by

\begin{equation}
\mathit{\Psi}=-\frac{1}{\mathcal{M}}\mathit{\Omega}^{3}\frac{\partial\mathit{\Pi}_{f}}{\partial\xi}\;\;\mathrm{for}\;\;\xi\in\left(\mathcal{A},\mathcal{A}+\gamma\right)\label{eq:scaled_poiseuille}
\end{equation}

\subsubsection*{Continuity}

The fluid continuity equation is given by

\begin{equation}
\frac{\partial\mathit{\Omega}}{\partial\tau}+\frac{1}{\xi^{d-1}}\frac{\partial\mathit{\check{\Psi}}}{\partial\xi}=0\label{eq:scaled_continuity_eq}
\end{equation}
in which the modified fluid flux is $\mathit{\check{\Psi}}=\xi^{d-1}\mathit{\Psi}$. 

\subsubsection*{Propagation criterion}

The \textit{k} asymptote \eqref{eq:k_asymptote} is scaled as

\begin{equation}
\mathit{\Omega}\sim\hat{\xi}^{1/2}\;\;\mathrm{for}\;\;\hat{\xi}\rightarrow0\label{eq:scaled_k_asymptote}
\end{equation}
where $\hat{\xi}=\mathcal{A}+\gamma-\xi$.

\subsubsection*{Boundary and initial conditions}

The boundary condition at fluid entrance is given by

\begin{equation}
\mathit{\check{\Psi}}\left(\xi,\tau\right)=\frac{1}{2\pi^{d-1}}\left(1-\frac{\text{d}\mathit{\Pi}_{w}}{\text{d}\tau}\right)\;\;\mathrm{at}\;\;\xi=\mathcal{A}\label{eq:scaled_inlet_bc}
\end{equation}
At the fracture tip, the boundary condition is scaled as

\begin{equation}
\mathit{\Omega}\left(\xi,\tau\right)=0,\;\mathit{\check{\Psi}}\left(\xi,\tau\right)=0\;\;\mathrm{at}\;\;\xi=\mathcal{A}+\gamma\label{eq:scaled_tip_bc}
\end{equation}
The initial condition is 
\begin{equation}
\mathit{\Pi}_{f}(\xi,\tau)=\bar{\sigma}_{d}^{\Delta T}(\xi,\tau^\prime)+\bar{\sigma}_{n}\left(\xi\right)+\mathit{\Pi}_{i},\;\;\;\left(\tau=0,\;\mathcal{A}<\xi<\mathcal{A}+\gamma_0\right)\label{eq:scaled_hf_ini_cond}
\end{equation}
in which $\tau^\prime=\left(t+t_c\right)/{t_\ast}$. At last, the scaled global volume balance equation is given by

\begin{equation}
2\int_{\mathcal{A}}^{\mathcal{A}+\gamma}\left(\pi\xi\right)^{d-1}\left(\mathit{\Omega}\left(\xi,\tau\right)-\mathit{\Omega}\left(\xi,0\right)\right)\text{d}\xi=\tau-\left(\mathit{\Pi}_w\left(\tau\right)-\mathit{\Pi}_w\left(0\right)\right)\label{eq:scaled_global_vol_bal}
\end{equation}

\subsection{Zero-viscosity solution}\label{subsec_inviscid_sol}

It is useful to investigate the case of uniform fluid pressure along the crack (or equivalently, the HF propagates in a toughness-dominated regime with dimensionless viscosity $\mathcal{M}\ll1$). For a HF with length $l$ under uniform fluid pressure, $p_f$, its mode I stress intensity factor can be computed as \citep{rice72,NiPr84,LeAb13}

\begin{equation}
\frac{K_\text{I}}{\sqrt{\pi l}}=\frac{2}{\pi}\int_{r_w}^{r_w+l_0}p(r)f\left(r,l,r_w\right)\frac{\text{d}r}{l\sqrt{1-\left(\frac{r-r_w}{l}\right)^2}}\label{eq:KI_k_regime}
\end{equation}
where
\[p\left(r\right)=p_f-\sigma^{\Delta T}\left(r\right)-\sigma_n(r)\]
\[f\left(x,l,r_w\right)=\left(\frac{r}{l+r_w}\right)^{d-1} \left[1+0.3\left(1-\frac{x-r_w}{l}\right)\left(\frac{r_w}{l+r_w}\right)^4\right]\]

Eq. \eqref{eq:KI_k_regime} can be used to determine the value of $p_f$ needed to initiate a HF by imposing $K_\text{I}=K_\text{IC}$, which can be further scaled as 

\begin{equation}
1=\frac{8}{\pi}\sqrt{2\gamma}\int_\mathcal{A}^{\mathcal{A}+\gamma}\mathit{\Pi}(\xi) \bar{f}\left(\xi,\gamma,\mathcal{A}\right)\frac{\text{d}\xi}{\gamma\sqrt{1-\left(\frac{\xi-\mathcal{A}}{\gamma}\right)^2}}\label{eq:scaled_KI_k_regime}
\end{equation}
with
\[\mathit{\Pi}\left(\xi\right)=\mathit{\Pi}_f-\bar{\sigma}^{\Delta T}\left(\xi\right)-\bar{\sigma}_n(\xi)\]
\[\bar{f}\left(\xi,\gamma,\mathcal{A}\right)=\left(\frac{\xi}{\gamma+\mathcal{A}}\right)^{d-1} \left[1+0.3\left(1-\frac{\xi-\mathcal{A}}{\gamma}\right)\left(\frac{\mathcal{A}}{\gamma+\mathcal{A}}\right)^4\right]\]

For a given initial length, $\gamma_0$, its corresponding breakdown pressure, $\mathit{\Pi}_b$, is then obtained by solving Eq. \eqref{eq:scaled_KI_k_regime}. The time of breakdown, $\tau_b$, can also be obtained by evaluating the gobal volume balance equation \eqref{eq:scaled_global_vol_bal}.

\subsection{Numerical scheme}

The numerical algorithm by \citet{LuGo17} is adopted to solve for the HF initiation and propagation. The elastic equation \eqref{eq:scaled_elasticity_axi} is discretized using a fixed grid of DD elements \citep{CrSt83}, and a 1-D finite difference scheme is applied for discretizing the Poiseuille law \eqref{eq:scaled_poiseuille} and continuity equation \eqref{eq:scaled_continuity_eq}. 

For the initiation phase, we fix the initial crack length $\gamma_0$ and apply a fixed time increment $\Delta\tau_\text{ini}$ for every time step until the propagation criterion \eqref{eq:scaled_k_asymptote} is satisfied. Then for the quasi-static HF growth, the solid–fluid coupled non-linear system is solved by an iterative scheme with two nested loops. During each step $i$, the fracture length is increased by a constant value, $\Delta\xi$, with $\gamma_i=\gamma_{i-1}+\Delta\xi$. The corresponding time step required to advance the fracture front by this increment, $\Delta\tau_i$, is solved as a part of the numerical solution. The discretization of the governing equations, as well as the length-controlled algorithm is discussed in detail in Appendix \ref{sec_Appendix_num_alg}. 

The main novelty of this work is the consideration of the thermo-elastic stress term, $\bar{\sigma}^{\Delta T}$, in the elasticity equation \eqref{eq:scaled_elasticity_axi}. In our numerical model, it is treated as an additional external load applied on the location of each DD element. In principle, this additional stress term evolves with time and needs to be solved together with the time step $\Delta\tau_i$. Nevertheless, since the time increment is much less than the total cooling duration (i.e., $\Delta t_i\ll t_i \ll t_c$ during fracture growth), the difference in the value of $\bar{\sigma}^{\Delta T}$ between two consecutive steps is negligible ($\bar{\sigma}^{\Delta T}\left(\tau_{i-1}\right)\approx\bar{\sigma}^{\Delta T}\left(\tau_i\right)$). Hence, we explicitly evaluate its value at previous step, $\bar{\sigma}^{\Delta T}\left(\tau_{i-1}\right)$ in the iterations to solve for the results of current step. Once the time increment $\Delta\tau_i$ is obtained, its value at every DD element is updated with the new cooling duration ($t_c+\tau_it_\ast$) and used in the following step.

\section{Results and Discussion}\label{sec_results}

In this section, our numerical model is first validated against (semi-)analytical solutions. Dimensional analysis is then performed to study the impact by various cooling and injection parameters. Finally, a hypothetical case study is carried out using realistic values under in-situ conditions at the Utah FORGE EGS project. In the numerical simulations, the cooling fluid is circulated in the wellbore to bring down the temperature at the borehole for certain amount of time, after which fluid is injected to create a HF. The effect of well cooldown on the pressure evolution and fracture geometry in the subsequent hydraulic fracturing stimulation is then examined.

\subsection{Validation}

\subsubsection{Thermo-elastic stress distribution}

We recall that the induced thermo-elastic stress disturbance in Eq. \eqref{eq:thermal_stress} is determined by two dimensionless parameters: dimensionless cooldown time $\mathcal{T}_c$ and thermal stress $\bar{\sigma}^{T_1}$. While the maximum magnitude of the stress is governed by $\bar{\sigma}^{T_1}$, its distribution over various distances from the wellbore wall is controlled by $\mathcal{T}_c$. Thermo-mechanical properties measured for Westerly granite \citep{McTi90,Chen16} will be used throughout this work to investigate the thermal conduction and its induced stress in typical granitic rocks in EGS. They are: $\kappa_T=1.09\times10^{-6}\;\text{m}^2/\text{s},\;\alpha_d=6\times10^{5}\;\text{N}\text{m}^{-2}\text{K}^{-1},\;\beta_d=2.4 \times 10^{-5}\;\text{K}^{-1},\;K=2.5 \times 10^{10}\;\text{Pa}$.

\begin{figure}[ht]
\begin{centering}
\includegraphics[width=.6\columnwidth]{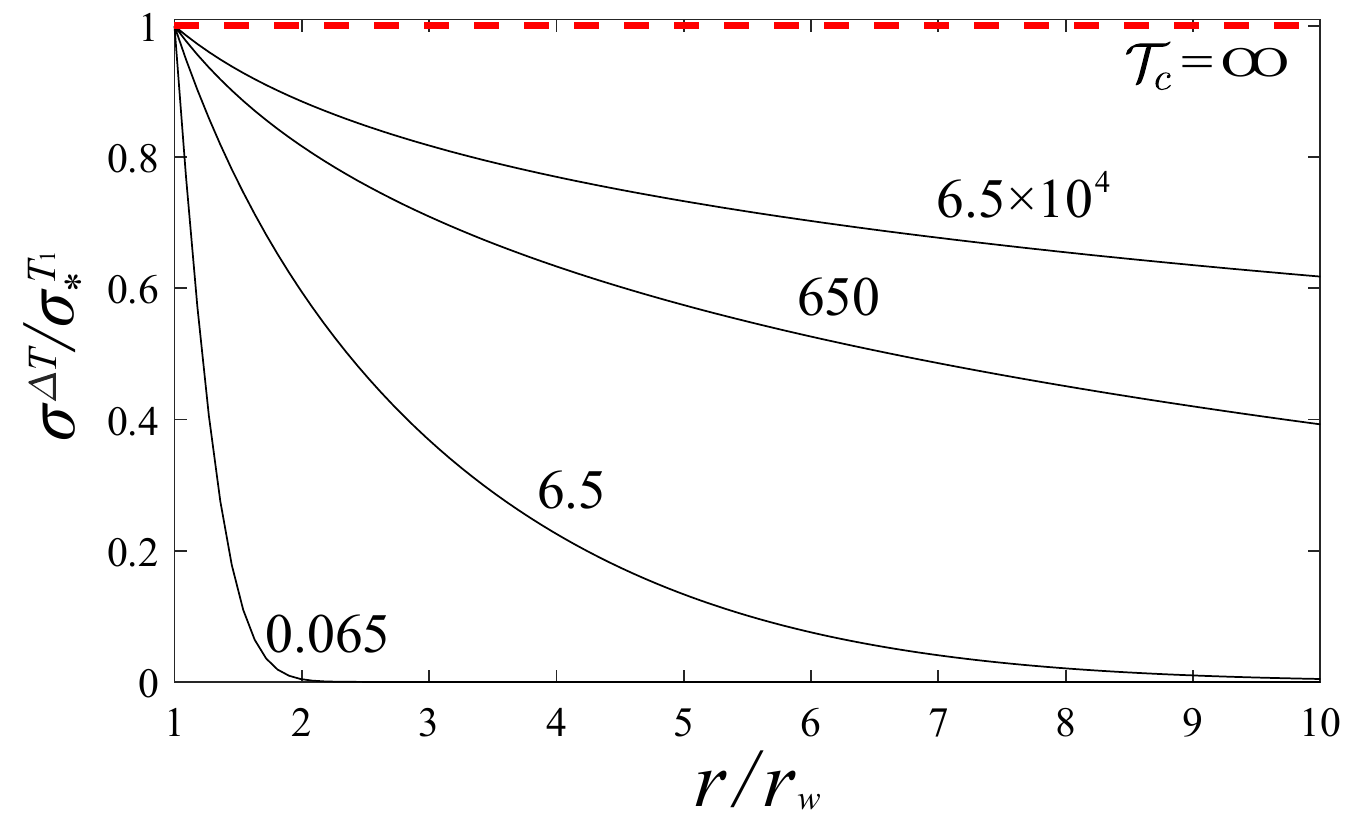}
\par\end{centering}
\caption{Distribution of thermally induced stress disturbance, normalized as $\sigma^{\Delta T}/\sigma_\ast^{T_1}$, at $\mathcal{T}_c=0.065,\;6.5,\;650,\;6.5\times10^{4}$ (corresponding to 600, 6$\times10^4$, 6$\times10^6$, 6$\times10^8$ seconds). Wellbore radius $r_w$ is set as 0.1 m. The red dashed line represents the constant temperature field in the limit of infinite cooling time $\mathcal{T}_c=\infty$.\label{fig:scaledsigt}}
\end{figure}

The temperature field (and its corresponding thermo-elastic stress given by Eq. \eqref{eq:thermal_stress}) is obtained using the inverse Laplace transform as described in Section \ref{subsec_heat_sol} and Appendix \ref{sec_Appendix_temp_sol}. The thermally induced stress field $\sigma^{\Delta T}$ (normalized by $\sigma_\ast^{T_1}$) at various dimensionless cooling time, $\mathcal{T}_c$, in the nearby region of the wellbore (up to 10$r_w$) is shown in \hyperref[fig:scaledsigt]{Fig.~\ref*{fig:scaledsigt}}. 

\subsubsection{Comparison with zero-viscosity solution}\label{subsubsec_l0_validation}

Our HF solver has been rigorously validated against existing analytical and numerical solutions in \citet{LuGo17,LuGo18} for HF growth in an isothermal environment. Here we compare the numerical results in limiting propagation regimes with the zero-viscosity solution under the influence of thermo-elastic stress.

Thanks to the scaling method, the physical processes taking place during fracture propagation are now reflected by one dimensionless quantity, namely the dimensionless viscosity $\mathcal{M}$. It describes the relative influence of the energy dissipated in viscous fluid flow in the fracture compared to the energy spent on the creation of new fracture surfaces \citep{Adac01,SaDe02,BuDe08,Deto16}. For $\mathcal{M}\ll1$, the HF is in a toughness-dominated (\textit{K}) regime, whereas a HF with $\mathcal{M}\geq1$ is growing in a viscosity-dominated (\textit{M}) regime. We will focus on the two cases in the limiting regimes (with $\mathcal{M}=0.001$ representing a typical HF growth in the toughness-dominated regime, and $\mathcal{M}=1$ for the HF propagation in the viscosity-dominated regime). Furthermore, the confining stresses ($\bar{\sigma}_h$,  $\bar{\sigma}_H$ and  $\bar{\sigma}_v$) are set to be zero in the following dimensional analysis. 

\begin{figure}[ht]
\begin{centering}
\includegraphics[width=.95\columnwidth]{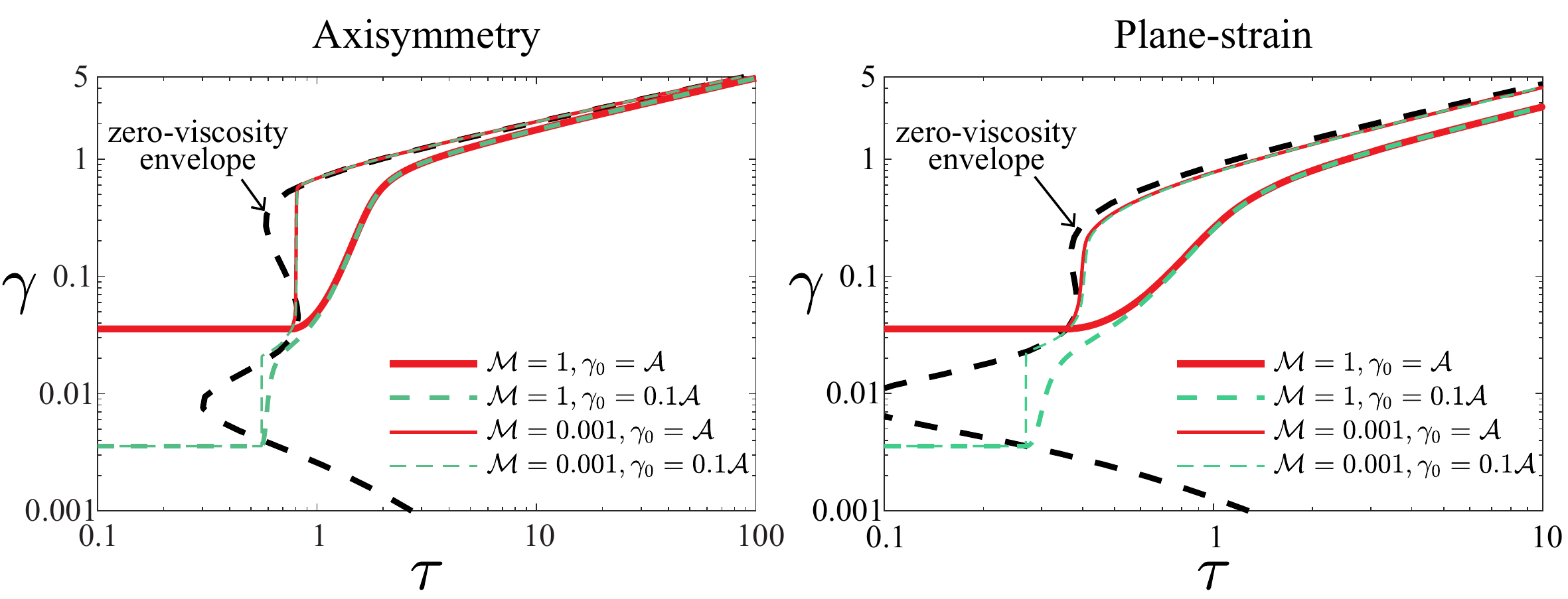}
\par\end{centering}
\caption{Evolution of dimensionless fracture length, $\gamma$, with time, $\tau$, for an axisymmetric (left) and a plane-strain (right) HF. For both geometries, the same wellbore cooling prior to the hydraulic fracturing treatment is considered, with $\mathcal{T}_c=0.065$ and $\bar{\sigma}^{T_1}=2.7$. Two HFs with initial crack length $\gamma_0=\mathcal{A}$ and $\gamma_0=0.1\mathcal{A}$ are studied under either the \textit{K} regime ($\mathcal{M}=0.001$) or the \textit{M} regime ($\mathcal{M}=1$). The zero-viscosity envelope is shown by dashed curves with unstable (lower) and stable (upper) branches.\label{fig:gmmavalidation}}
\end{figure}

The method presented in Section \ref{subsec_inviscid_sol} allows us to solve for the HF breakdown time, $\tau_b$, as well as its corresponding breakdown pressure, $\mathit{\Pi}_b$, for a uniformly pressurized HF with given initial crack length, $\gamma_0$. An envelope with both upper and lower bounds is recovered when plotting such zero-viscosity solutions for the length evolution with time in \hyperref[fig:gmmavalidation]{Fig.~\ref*{fig:gmmavalidation}} (given the pre-injection wellbore cooldown for 10 minutes with 30 $^\circ$C cooling at the borehole, corresponding to $\mathcal{T}_c=0.065$ and $\bar{\sigma}^{T_1}=2.7$). Simulation results of two HFs with long and short initial cracks ($\gamma_0=\mathcal{A}$ and $\gamma_0=0.1\mathcal{A}$) are compared with the zero-viscosity solutions under both axisymmetry and plane strain conditions in \hyperref[fig:gmmavalidation]{Fig.~\ref*{fig:gmmavalidation}}. 

It is seen that the length of all starter cracks remains constant with time until they hit the inviscid solution curve, indicating the HF initiation at time, $\tau_b$. For HFs in the toughness-dominated regime ($\mathcal{M}=0.001$), an instability in length is observed as its value jumps instantly to the upper branch of the zero-viscosity curve, followed by a stable growth along the envelope until it gets a second opportunity to jump again to a higher upper branch (two jumps in fracture length are detected in the case of a shorter initial crack $\gamma_0=0.1\mathcal{A}$). For a viscosity-dominated HF ($\mathcal{M}=1$), the fracture growth is slowed down (no sudden jump in length evolution) due to the viscous dissipation. Convergence to the upper limit of the envelope is seen at large time in the axisymmetric configuration. Such finding is consistent with previous studies on the transition from \textit{M} regime at early time to \textit{K} regime in late time propagation for an axisymmetric (or penny-shaped) HF \citep{SaDe02,AbLe13,Deto16}.

\begin{figure}[ht]
\begin{centering}
\includegraphics[width=.95\columnwidth]{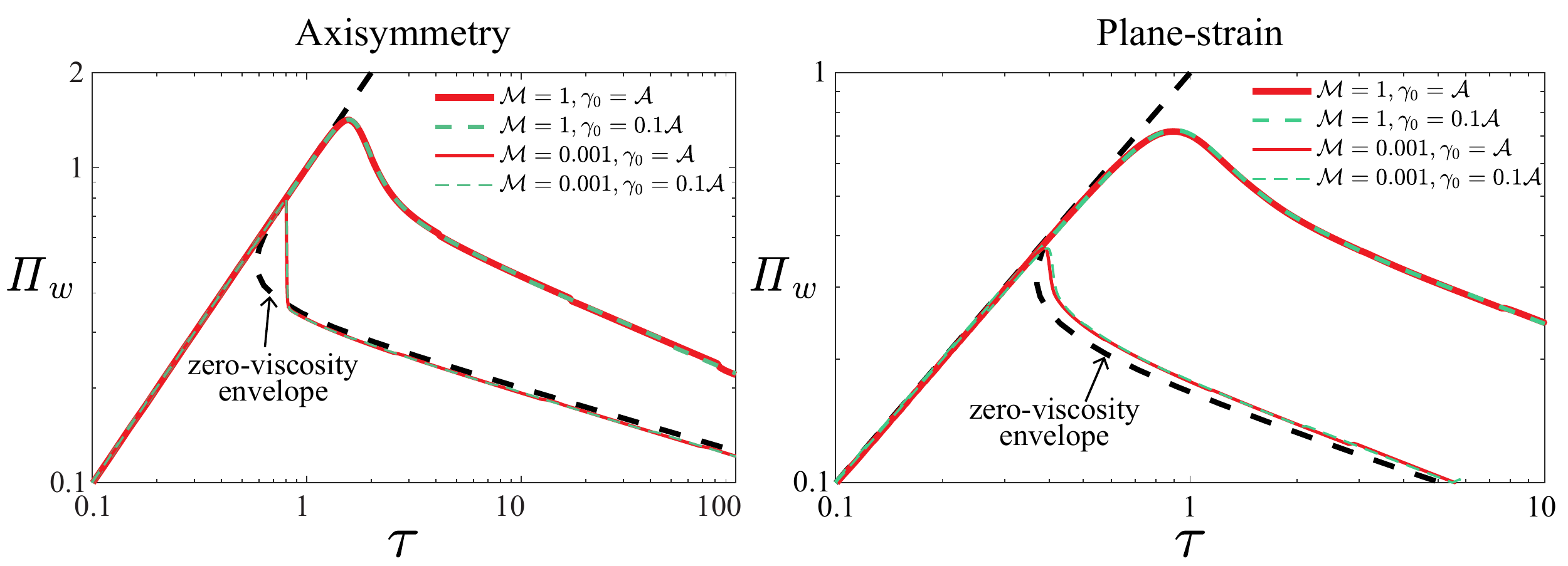}
\par\end{centering}
\caption{Evolution of dimensionless wellbore pressure, $\mathit{\Pi}_w$, with time, $\tau$, for an axisymmetric (left) and a plane-strain (right) HF under the same conditions as in \hyperref[fig:gmmavalidation]{Fig.~\ref*{fig:gmmavalidation}}.\label{fig:piwvalidation}}
\end{figure}

Similar behaviors are exhibited in the evolution of wellbore pressure, $\mathit{\Pi}_w$ (\hyperref[fig:piwvalidation]{Fig.~\ref*{fig:piwvalidation}}). In the \textit{K} regime, the wellbore pressure drops drastically once HF initiation takes place, whereas it keeps increasing for a while after fracture starts growing until it reaches its peak value in the \textit{M} regime. These results imply that the breakdown pressure ($\mathit{\Pi}_b$), defined as the peak value of the wellbore pressure, depends on the dimensionless viscosity. $\mathit{\Pi}_w$ converges to the lower branch of the envelope when stable HF growth is achieved.

Surprisingly, the breakdown pressure is found to be independent of the initial crack length, $\gamma_0$ - the wellbore pressure evolution remains the same regardless of the variation in $\gamma_0$ under either regime ($\mathcal{M}=0.001$ or $\mathcal{M}=1$). It is possibly due to strong thermo-elastic stress concentration near the wellbore - after the preceding cooling circulation, the thermally induced stress front has extended to the distance of one wellbore radius $r_w$ away from the borehole wall (see stress distribution of the case $\mathcal{T}_c=0.065$ in \hyperref[fig:scaledsigt]{Fig.~\ref*{fig:scaledsigt}}). Consequently, both long and short starter cracks are under the influence of the thermo-elastic stress, which is evidently a primary determinant of the breakdown pressure. Since the realistic range for the initial flaw length is $l_0\in\left(0.01r_w,r_w\right)$ \citep{LeAb13}, and the cooling operation at the EGS project most likely lasts longer than 10 minutes ($\mathcal{T}_c=0.065$ in \hyperref[fig:scaledsigt]{Fig.~\ref*{fig:scaledsigt}}), we conclude that the selection of $l_0$ should not affect the breakdown pressure when the thermally induced stress is present. Note that, the defect length $l_0$ is commonly considered as a material-like parameter, which can be represented by the so-called Irwin's material lengthscale $l_m=K_\text{IC}^2/\sigma_T^2$ \citep{Leca12}. Based on measured mechanical properties of granitic rocks in laboratory experiments \citep{Mart93,NaTa09,KeBu14,LuUw15,FeLu16,WiLu18,LuGo20,LuZh22}, the ratio of the mechanical properties, $K_\text{IC}^2/\sigma_T^2$, ranges from $1/4$ to $1/10$ in most cases, leading to a lengthscale of $l_m\sim O\left(10^{-2}\right)\text{m}$. Consequently, $l_m$ has the same order of magnitude as the wellbore radius, $r_w$. From now on, all simulations will be conducted using a fixed initial flaw length that equals the wellbore radius, $l_0=r_w$.

\subsection{Impact of the cooling time}

\begin{figure}[ht]
\begin{centering}
\includegraphics[width=\columnwidth]{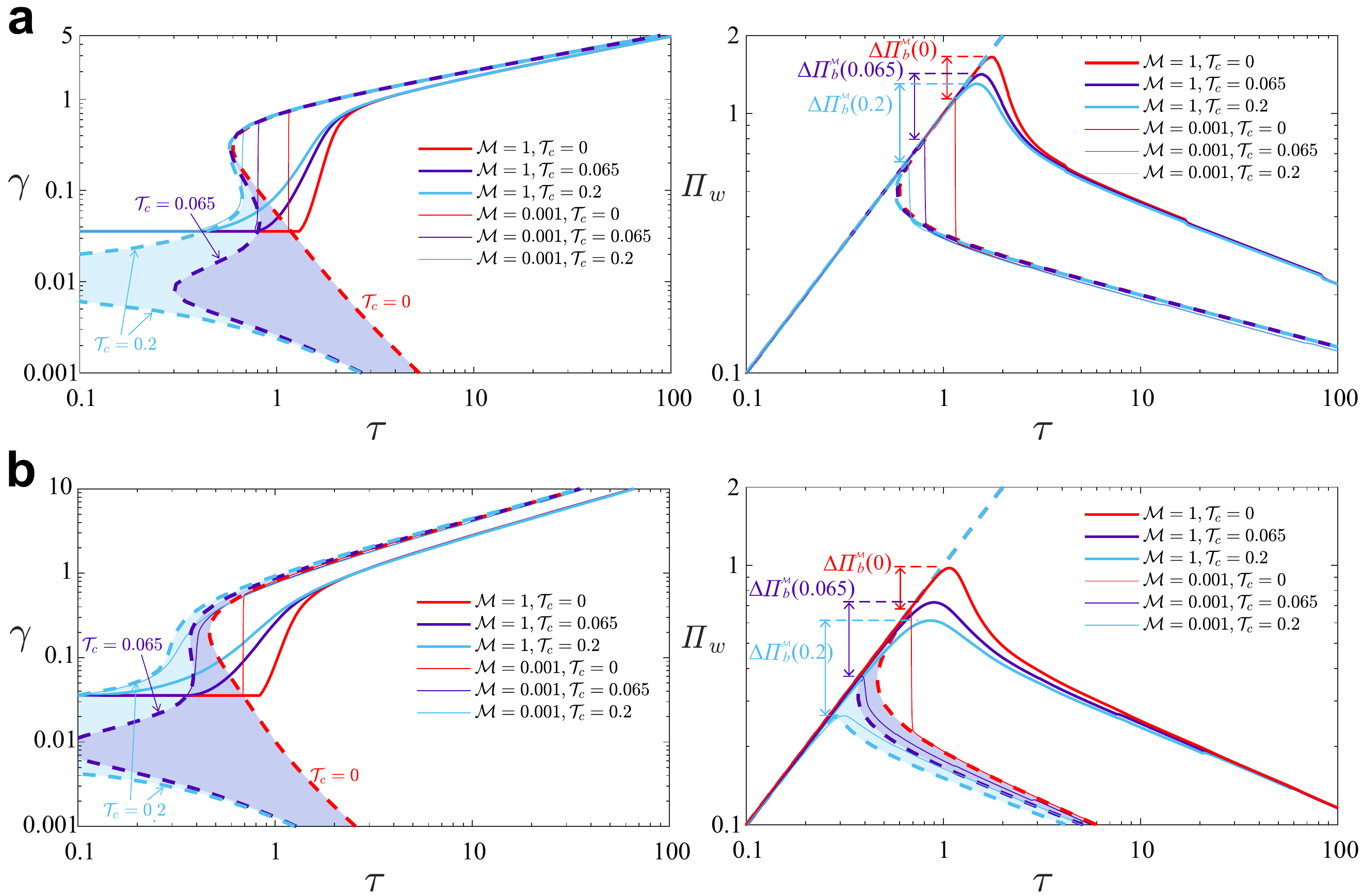}
\par\end{centering}
\caption{Evolution of dimensionless fracture length, $\gamma$, and the wellbore pressure, $\mathit{\Pi}_w$, with time, $\tau$, for an axisymmetric (a) and a plan-strain (b) HF with different fluid circulation times ($\mathcal{T}_c=0,\;0.065,\;0.2$). For all cases, the characteristic stress, $\bar{\sigma}^{T_1}$, is set to be 2.7, which is caused by a constant temperature drop at the wellbore of 30 $^\circ$C. \label{fig:varytc}}
\end{figure}

A parametric study is carried out to examine the impact of the cooldown process. In this section, identical fluid injection and confining stresses conditions described in Section \ref{subsubsec_l0_validation} are applied, rendering results of HF growth in the \textit{K} and \textit{M} regimes, respectively. We first explore the HF behavior after different pre-injection circulation times, $\mathcal{T}_c$. Three values of $\mathcal{T}_c$ (0, 0.065 and 0.2, corresponding to no cooling, 10 minutes, and 30 minutes cooling circulations) are utilized in the simulations (subjected to the $T_1=30\;^\circ$C constant temperature difference boundary condition equivalent to $\bar{\sigma}^{T_1}=2.7$). The HF initiation and propagation following various duration of wellbore cooldown are plotted (fracture length and wellbore pressure evolution) along with their respective inviscid solution curves in \hyperref[fig:varytc]{Fig.~\ref*{fig:varytc}}. Comparisons among different cases lead to several observations: 

\begin{enumerate}
\item In the presence of cooling circulation, the lower branch of the zero-viscosity envelop undergoes a leftward shift in the fracture length plot (see the differences between the cooling and no cooling envelops highlighted in \hyperref[fig:varytc]{Fig.~\ref*{fig:varytc}}), indicating an earlier HF initiation caused by the near-wellbore thermo-elastic stress distribution. For both axisymmetry and plane-strain geometries, the promotion in HF initiation is most obvious for starter cracks with intermediate values of $\gamma_0$ around 0.01. 
\item The breakdown pressure envelope's lower bound also shifts downward in the plane-strain configuration due to cooling, which implies that cooling also reduces the pressure needed for stable fracture propagation. Interestingly, the lower bound in the axisymmetric case stays unaffected by the cooling, with all curves overlapping each other in spite of the different cooling times.
\item For HFs in both regimes, cooling circulation lowers the breakdown pressure. Longer duration of cooldown results in more significant decline of $\mathit{\Pi}_b$. 
\item In general, $\mathit{\Pi}_b$ in the viscosity-dominated (\textit{M}) regime is larger than its value the toughness-dominated (\textit{K}) regime. Without cooling, the drop in $\mathit{\Pi}_b$ from the \textit{M} regime ($\mathcal{M}=1$) to \textit{K} regime ($\mathcal{M}=0.001$), $\Delta\mathit{\Pi}_b^\mathcal{M}\left(\mathcal{T}_c=0\right)$ (see \hyperref[fig:varytc]{Fig.~\ref*{fig:varytc}}), is 30.73\% for axisymmetry and 30.53\% for plane-strain. After wellbore cooldown, it is increased to $\Delta\mathit{\Pi}_b^\mathcal{M}\left(\mathcal{T}_c=0.065\right)=43.88\%,\;\Delta\mathit{\Pi}_b^\mathcal{M}\left(\mathcal{T}_c=0.2\right)=50.02\%$ (axisymmetry), and $\Delta\mathit{\Pi}_b^\mathcal{M}\left(\mathcal{T}_c=0.065\right)=48.1\%,\;\Delta\mathit{\Pi}_b^\mathcal{M}\left(\mathcal{T}_c=0.2\right)=57.04\%$ (plane-strain). Thus, more intensive cooling amplifies the reduction of $\mathit{\Pi}_b$ when lowering the dimensionless viscosity $\mathcal{M}$ from \textit{M} to \textit{K} regimes.
\end{enumerate}

\subsection{Effect of the cooling temperature}

\begin{figure}[ht]
\begin{centering}
\includegraphics[width=\columnwidth]{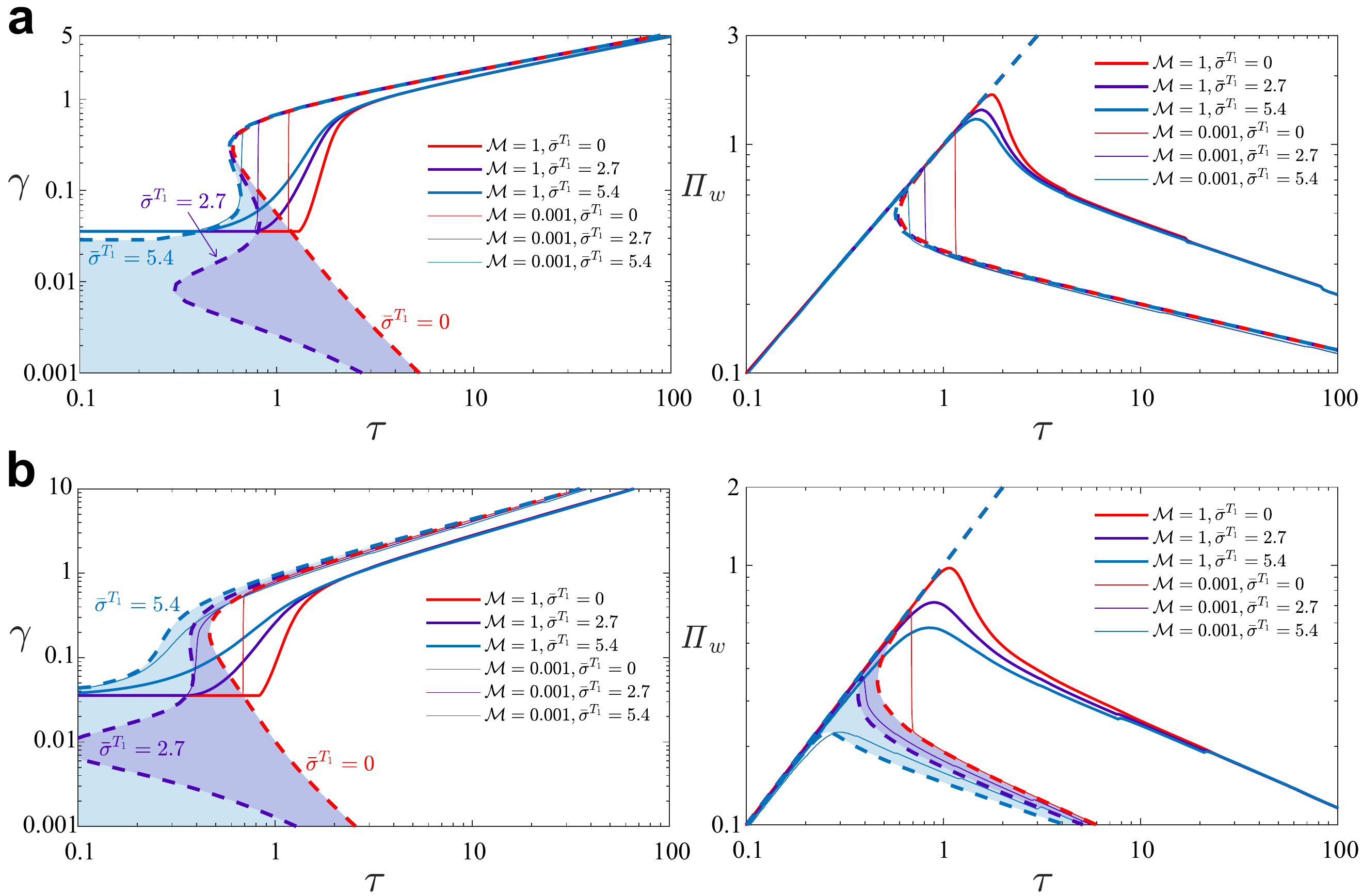}
\par\end{centering}
\caption{Evolution of dimensionless fracture length, $\gamma$, and the wellbore pressure, $\mathit{\Pi}_w$, with time, $\tau$, for an axisymmetric (a) and a plan-strain (b) HF under different thermally induced stresses ($\bar{\sigma}^{T_1}=0,\;2.7,\;5.4$) with a constant cooling time $\mathcal{T}_c=0.065$. \label{fig:varysigt}}
\end{figure}

The influence of temperature difference, $T_1$, and its associated characteristic thermo-elastic stress, $\bar{\sigma}^{T_1}$, is reported in \hyperref[fig:varysigt]{Fig.~\ref*{fig:varysigt}}. Results obtained for three cases of $\bar{\sigma}^{T_1}=0,\;2.7,\;5.4$ (generated by $T_1=0,\;30,\;60\;^\circ$C) with the cooling duration fixed at $\mathcal{T}_c=0.065$ are plotted. Similar trend is shown with increasing $\bar{\sigma}^{T_1}$. Larger $\bar{\sigma}^{T_1}$ results in HF earlier initiation and lower breakdown pressure in both geometries. Notably, a thermally induced stress of $\bar{\sigma}^{T_1}=5.4$ is sufficiently large to initiate any starter defect with length $\gamma_0<0.03$ (about the same size as the wellbore radius). 

Again, we compare the reduction in breakdown pressure from the \textit{M} to the \textit{K} limit under different thermo-elastic stresses:

\textbf{Axisymmetry} - $\Delta\mathit{\Pi}_b^\mathcal{M}\left(\bar{\sigma}^{T_1}=0\right)=30.73\%$, $\;\Delta\mathit{\Pi}_b^\mathcal{M}\left(\bar{\sigma}^{T_1}=2.7\right)=43.88\%$, $\;\Delta\mathit{\Pi}_b^\mathcal{M}\left(\bar{\sigma}^{T_1}=5.4\right)=50.11\%$.

\textbf{Plane-strain} -
$\Delta\mathit{\Pi}_b^\mathcal{M}\left(\bar{\sigma}^{T_1}=0\right)=30.53\%$, $\;\Delta\mathit{\Pi}_b^\mathcal{M}\left(\bar{\sigma}^{T_1}=2.7\right)=48.1\%$, $\;\Delta\mathit{\Pi}_b^\mathcal{M}\left(\bar{\sigma}^{T_1}=5.4\right)=60.4\%$.

The drop in $\mathit{\Pi}_b$ is found to be similar in both geometries, with slightly larger values obtained in the plane-strain HF. 

\subsection{Transition from longitudinal to transverse fracture geometry}

The competition between the transverse and longitudinal fracture geometries plays a critical role in the determination of pressure evolution and the formed HF pattern in near-wellbore regions \citep{WeDe94,Weij95,LeAb13,BeDj21}. By adopting the same method used by \citet{LeAb13}, we investigate the HF orientation during a hydraulic fracturing treatment under realistic fluid circulation and injection conditions. More specifically, material properties and in-situ stress conditions at the Utah FORGE EGS site \citep{MoMc20,XiDa21} are used in this study. These parameters are listed in \hyperref[tab:forgenumbers]{Table.~\ref*{tab:forgenumbers}}. The hydraulic fracture (either a transverse or longitudinal) is set to be initiated from a horizontal wellbore drilled along the direction of the minimum in-situ stress, $\sigma_h$. The intermediate and maximum in-situ stresses are set to be  $\sigma_H$ and  $\sigma_v$, respectively. Therefore, the transverse axisymmetric HF propagates normal to $\sigma_h$ as demonstrated in \hyperref[fig:setup]{Fig.~\ref*{fig:setup}c}, and the longitudinal bi-wing HF grows on the vertical plane normal to $\sigma_H$ (\hyperref[fig:setup]{Fig.~\ref*{fig:setup}d}). To determine the orientation of the HF, we quantify the breakdown pressure, $P_b$, required to propagate the HF of both types 
and whichever type needs a lower $P_b$ becomes the preferred geometry. Alternatively, the preferred HF geometry can be decided by comparing the wellbore pressure, $P_w$, throughout the propagation of the HF (i.e., comparing the evolution of $P_w$ with fracture length $l$ in both geometries). This approach is also used in \citet{LeAb13}. Since both methods (comparison of $P_b$, and comparison of $P_w$) lead to the same conclusions in our study, we will adopt the former approach in the following analysis, and concentrate on the interpretation of the peak wellbore pressure in different scenarios.

\begin{table}[ht]
\caption{Fluid injection and in-situ conditions utilized in the numerical simulations to predict the wellbore pressure and geometry of generated HF from a horizontal well at an EGS site.} \label{tab:forgenumbers}
\begin{tabular}{ll}
\toprule%
Fracture toughness $K_\text{IC}$ (MPa$\sqrt{\text{m}}$) & 3 \\
Young's modulus $E$ (GPa) & 51.3 \\
Poisson's ratio $\nu$ & 0.26 \\
Heat diffusivity coefficient $\kappa_T$ (m$^2/$s) & $1.09\times10^{-6}$ \\
Drained thermo-elastic effective stress coefficient $\alpha_d$ ($\text{N}\text{m}^{-2}\text{K}^{-1}$) & $6\times10^{5}$ \\
Fluid viscosity $\mu$ (Pa$\cdot$s) & 0.002 \\
Injection rate $Q_0$ (m$^3/$s) & 0.0132 \\
Wellbore radius $r_w$ (m) & 0.1 \\
Wellbore interval length $L_w$ (m) & 10 \\
Compressibility $U$ (m$^3/$Pa) & 2$\times10^{-10}$ \\
Maximum in-situ stress $\sigma_v$ (MPa) &  65.5 \\
Intermediate in-situ stress $\sigma_H$ (MPa) & 51 \\
Minimum in-situ stress $\sigma_h$ (MPa) & 44.7 \\
\botrule
\end{tabular}
\end{table}

\begin{figure}[ht]
\begin{centering}
\includegraphics[width=.9\columnwidth]{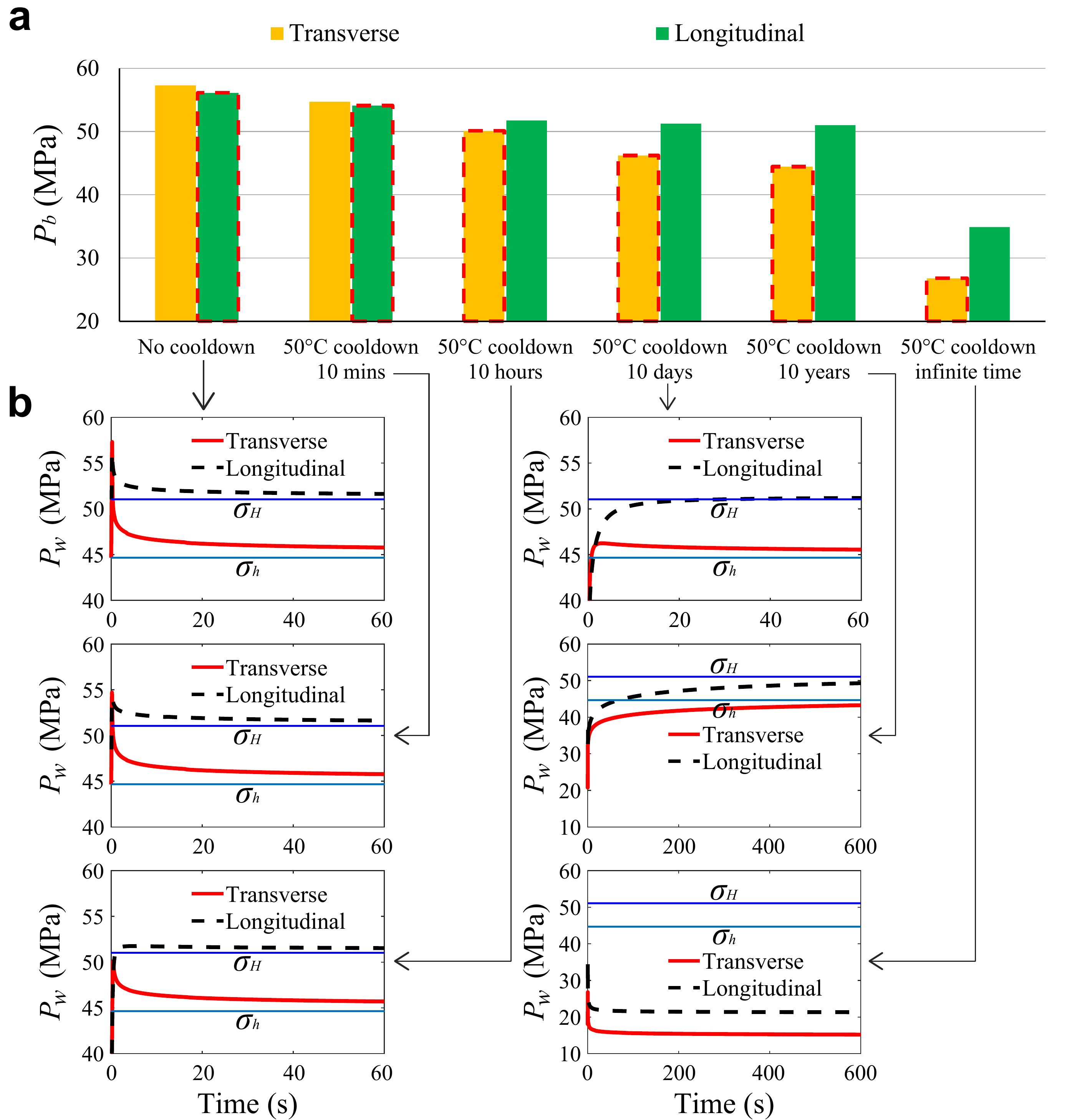}
\par\end{centering}
\caption{a, Peak wellbore pressure, $P_b$, of hydraulic fracturing following various cooling times (No cooling, 10 minutes, 10 hours, 10 days, 10 years, and infinite time) with a constant temperature difference (50 $^\circ$C) at the wellbore. The preferred fracture geometry in each case is highlighted by red dashed lines. b, Wellbore pressure versus injection time for both geometries in all cases. \label{fig:pb50c}}
\end{figure}

A series of simulations are performed for the same hydraulic fracturing activity (\hyperref[tab:forgenumbers]{Table.~\ref*{tab:forgenumbers}}) with distinct cooling parameters. We first investigate the case of a 50 $^\circ$C cooldown at the borehole. A wide range of antecedent circulation time is studied (10 minutes, 10 hours, 10 days, 10 years, and infinite cooling time, of which the solutions are derived in Appendix \ref{sec_Appendix_temp_sol}). The wellbore pressure evolution during hydraulic fracturing is plotted in \hyperref[fig:pb50c]{Fig.~\ref*{fig:pb50c}b}. The breakdown pressure, $P_b$, of both geometries are obtained and the preferred geometry is decided. Overall, the breakdown pressure decreases with an increasing cooling time in both geometries. However, such decreasing trend over time differs depending on the configuration. As a result, a clear transition in the favorable geometry from the longitudinal to the transverse orientation is recovered with longer cooling times (For cooling duration longer than 10 hours, the transverse geometry needs lower $P_b$ compared to the longitudinal configuration).

Another notable observation is the absence of a discernible peak value in $P_w$ after a long term cooling, which would otherwise be considered a common feature associated with HF breakdown in an isothermal environment (see the leftmost case in \hyperref[fig:pb50c]{Fig.~\ref*{fig:pb50c}b} with apparent peaks, compared to the case of 10-year cooldown, in which $P_w$ keeps increasing throughout the plotted period). The peak disappears firstly in the longitudinal configuration (after 10 hours' cooldown), and it vanishes after 10 days of cooling in the transverse geometry. Such alteration in the pressure evolution apparently results from the thermo-elastic stress concentration near the wellbore. 

Generally, as the HFs grow longer, the fluid pressure that drives its propagation should converge to the far-field stress normal to the fracture faces, which in the above mentioned wellbore configuration is either $\sigma_h$ acting on the transverse HF, or $\sigma_H$ applied normal to the longitudinal direction. It is seen in the cases with shorter cooling times (up to 10 days) that $P_w$ in both geometries rapidly converges to its respective far-field stress ($\sigma_h$ or $\sigma_H$). Nonetheless, it takes much longer time ($\sim$600 s) for $P_w$ to approach the values of $\sigma_h$ and $\sigma_H$ after a 10-year cooling period. Furthermore, in the limiting case of infinite cooling time, the stress field is greatly impaired, causing the pressure to be stabilized at a significantly lower level.

\begin{figure}[ht]
\begin{centering}
\includegraphics[width=.5\columnwidth]{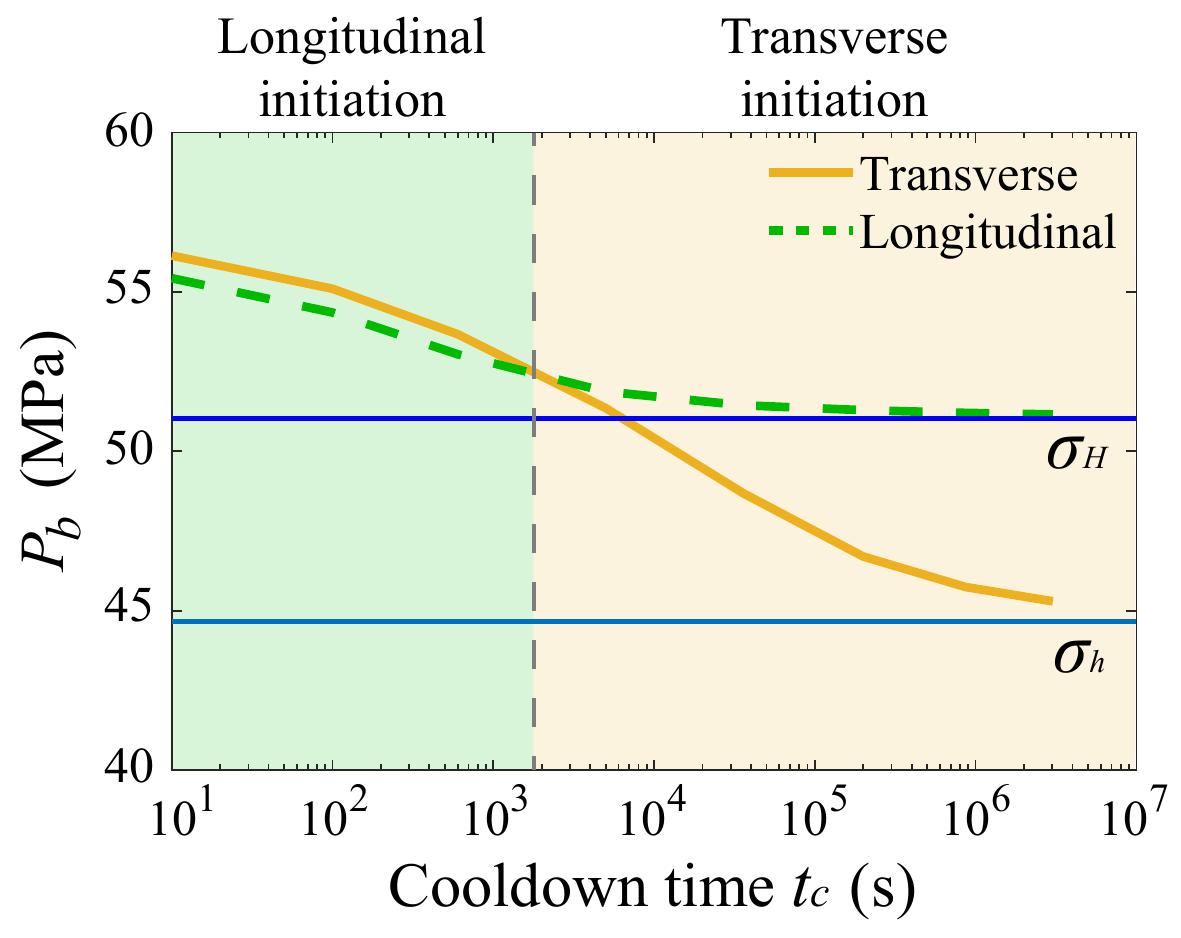}
\par\end{centering}
\caption{Transition from longitudinal HF initiation with shorter wellbore cooldown times to transverse HF initiation in cases of longer cooling periods takes place at $\sim$2000 s. The cooling temperature is set to be 100 $^\circ$C. \label{fig:pb100c}}
\end{figure}

Plotting the relationship between breakdown pressure, $P_b$, and cooldown time, $t_c$, enables us to explore the sensitivity of $P_b$ in response to varying $t_c$ in both geometries (see \hyperref[fig:pb100c]{Fig.~\ref*{fig:pb100c}} for a constant cooling temperature at $T_1=100\;^\circ$C). When the thermo-elastic stress is insignificant (short-term cooling), $P_b$ is determined mainly by near-wellbore stress concentration and fluid injection conditions, which favors the longitudinal initiation. The different values of $P_b$ arise from the discrepancy in the total energy spent on both (1) creation of new fracture surfaces (\textit{K}), and (2) dissipation of viscous fluid (\textit{M}) in two fracture geometries. In the case of an insignificant thermal effect, the viscous dissipation requires more energy for the transverse HF compared to the longitudinal geometry, thereby resulting in a higher energy input, $P_bQ_0$ \citep{LeAb13}.

With an increasing $t_c$, the cooling induced tensile thermo-elastic stress substantially reduces the total compressive stress acting on the crack, causing $P_b$ to approach the magnitude of the far-field stresses, i.e., $P_b\rightarrow\sigma_h$ (transverse) and $P_b\rightarrow\sigma_H$ (longitudinal). Since $\sigma_H>\sigma_h$, the transverse fracture configuration becomes dominant in the long term. It is also worthwhile to note that a clear peak value in $P_w$ vanishes with long term cooling as illustrated in \hyperref[fig:pb50c]{Fig.~\ref*{fig:pb50c}}. Hence, $P_b$ for large cooling period should merge with the respective far-field stress when the HF outgrows the thermal conduction zone.

Additionally, \hyperref[fig:pb100c]{Fig.~\ref*{fig:pb100c}} implies that bigger cooling temperature also accelerates the transition in HF geometry, as shorter cooling time is needed for the transition to take place with $T_1=100\;^\circ$C ($\sim$2000 s of circulation), compared to $\sim$5000 s in the case of 50 $^\circ$C cooldown.

\subsection{Implications for hydraulic fracturing field applications}

The results from this numerical study have important implications for field practices. In in-situ stress testing, the minimum horizontal stress, $\sigma_{h}$, is usually estimated by the shut-in pressure $P_s$ with $\sigma_{h}=P_s$ \citep{CoVa84,DeWa89,HaCo03}. The pre-requisite to this classical analysis is that the HF is oriented so that $\sigma_h$ is acting normal to the HF plane. This assumption is most likely to be satisfied in a situation with a vertical well drilled into a formation where $\sigma_h<\sigma_v$, where $\sigma_v$ is the vertical stress. However, if a well is drilled parallel to $\sigma_h$ (i.e. because of a desire to stimulate with transverse fractures), a longitudinal initiation direction can make the interpretation of minimum stress misleading as such a HF will not be perpendicular to $\sigma_h$ direction. In this scenario, a transverse initiation can be promoted by cooling and when this is the case, the shut-in pressure is more accurately interpreted to correspond to $\sigma_h$. However, the transverse initiation makes it inappropriate to take the next step in classical interpretation wherein the breakdown and/or reopening pressure are analyzed based on the solution for a plane strain wellbore to estimate the maximum stress \citep{HaCo03}. A similar scenario involving longitudinal versus transverse initiation and its impacts on mini-HF growth and subsequent interpretation is also expected for a vertical well in reverse/thrust stress regimes where $\sigma_v<\sigma_h$. The key point is that wellbore cooling can lead to transverse initiation when the well is drilled parallel to the least compressive stress, and this fact must be included in the planning, execution, and interpretation of mini-HF testing. 

For hydraulic fracturing stimulations, the cooling process can lower the breakdown pressure, making it easier to initiate the HFs. Additionally, HFs tend to propagate in the direction normal to $\sigma_h$ when growing far away from the wellbore. For a horizontal well drilled in the minimium horizontal stress direction, in the absence of cooling it is possible, perhaps even most favorable in most cases for the HF to start with the longitudinal geometry. As it grows, it will likely re-orient to be aligned with the transverse plane. However, this reorientation can cause tortuosity, which in turn can lead to high injection pressures, impediments to proppant transport, and high resistance to production (in the case of petroleum wells) or circulation (in the case of EGS wells) \citep{WaBa22}. Wellbore cooling can potentially prevent such non-planar growth of the HF in the near field by promoting the transverse initiation.

\section{Conclusions}\label{sec_conclusions}

A numerical model is presented for studying the initiation and propagation of either a plane-strain or an axisymmetric hydraulic fracture initiating and emanating from a circular borehole. The novelty of this study is the accounting for the thermo-mechanical effect due to cooling circulation at the wellbore prior to the hydraulic fracturing treatment. The impact of various cooling parameters on the fracture behavior has been studied using this model. A case study is carried out, using realistic parameters inspired by the Utah FORGE EGS site, to predict possible scenarios of the pressure evolution and fracture geometry. 

Based on our simulation results, three main conclusions can be drawn. First, cooling circulation contributes to an earlier fracture initiation and a decrease in the hydraulic fracturing breakdown pressure. Longer duration and lower circulation temperature results in a greater reduction of the breakdown pressure. By targeting a toughness-dominated hydraulic fracture and employing an extended circulation period with cold fluid, the lowest possible breakdown pressure can be achieved.

Second, the thermo-elastic stress exerts substantial impact on the wellbore pressure evolution, which delays the convergence to the far-field stress. Under the extreme condition of infinitely large cooling duration, the wellbore pressure evolution is drastically altered and it drops to a much lower level below the respective far-field stress.

Finally, in the absence of cooling, it is already known that longitudinal hydraulic fracture initiation can be favored even in cases where the well is drilled parallel to the least compressive stress (i.e. a horizontal well drilled in the miniminum horizontal stress direction). Here we have shown that cooling of sufficient magnitude and for a sufficient duration can lead to favoring of transverse initiation. This can profoundly impact the planning and interpretation of mini-HF tests for determining in-situ stresses. It also has the potential to have a mitigating effect on unwanted near-wellbore tortuousity that can be encountered during stimulation treatments when longitudinally-initiating HFs rotate as they grow away from the wellbore for a transverse alignment in the far field. With sufficient cooling, the initiation can also be transverse, which is expected to lead to a simpler fracture geometry overall. Because of these important impacts on HF initiation and early growth, when cooling is used to protect equipment performance in mini-HF and/or when it is inevitable as a part of stimulation HF, its effects should be considered for both planning the treatment and interpretation of the results. 

\section*{Acknowledgments}

This work was performed at the University of Pittsburgh with funding provided by DOE EERE Geothermal Technologies Office to Utah FORGE and the University of Utah under Project DE-EE0007080 Enhanced Geothermal System Concept Testing and Development at the Milford City, Utah Frontier Observatory for Research in Geothermal Energy (Utah FORGE) site. Support was via Subcontract No. 845391 to Battelle Memorial Institute for Utah FORGE Project 2439. Additional support for APB is provided by the RK Mellon Faculty Fellowship in Energy.

\begin{appendices}

\section{Solution to 1-D axisymmetric heat diffusion equation \label{sec_Appendix_temp_sol}}

After applying Laplace transformation to Eq. \eqref{eq:diffusion_eq}, we obtain \citep{Chen16}

\begin{equation}
\frac{1}{r}\frac{\partial}{\partial r}(r\frac{\partial \widetilde{T}}{\partial r})-\frac{s}{c_a}\widetilde{T}=0\label{eq:laplace_diffusion_eq}
\end{equation}
where $s$ is the Laplace transform parameter, $\widetilde{T}$ is defined as the Laplace transform of $T$, and $c_a$ represents the entropy diffusivity coefficient. The solution to Eq. \eqref{eq:laplace_diffusion_eq} subject to boundary conditions \eqref{eq:diff_ini_cond} is found as

\[\widetilde{T}=\frac{T_1 \text{K}_0(r\sqrt{s/\kappa_T})}{s \text{K}_0(r_0\sqrt{s/\kappa_T})}\]
in which $\text{K}_0$ is the modified Bessel function of the second kind of order 0. A numerical inversion technique \citep{Steh70} is then applied to find the temperature in time. Here we briefly introduce this method. Letting $\tilde{f}(s)$ be the transform of $f(t)$, the solution in time is approximated by 

\begin{equation}
f(t)\approx\frac{\text{ln2}}{t}\sum_{n=1}^N C_n \tilde{f}\left(n\frac{\text{ln2}}{t}\right)\label{eq:stehfest_method}
\end{equation}
with the coefficient $C_n$ expressed as
\[C_n=\left(-1\right)^{n+N/2}\sum_{k=\left(n+1\right)/2}^{\text{min}\left(n,N/2\right)}\frac{k^{N/2}(2k)!}{(N/2-k)!k!(k-1)!(n-k)!(2k-n)!} \]

Additionally, Laplace transform has the following correlation:

\[\lim_{t\to\infty}f(t)=\lim_{s\to0}s\tilde{f}(s)\]

Consequently, the solution for temperature at infinite time can be derived as

\begin{equation}
\lim_{t\to\infty}T(t,r)=\lim_{s\to0}s\widetilde{T}(s)= T_1\lim_{s\to0}\frac{\text{K}_0(r\sqrt{s/\kappa_T})}{\text{K}_0(r_0\sqrt{s/\kappa_T})}=T_1
\end{equation}

\section{Numerical algorithm}\label{sec_Appendix_num_alg}

\subsection{Discretization}

The elasticity equation \eqref{eq:scaled_elasticity_axi} is discretized into a linear system using the DD method. For an axisymmetric HF, The crack is discretized into a mesh of $m$ piece-wise constant-strength DD elements. In the plane-strain geometry, additional elements are used to discretize the circular wellbore wall. The crack width at the $i$th element, $\mathit{\Omega}_{i}\;\left(i=1,...,m\right)$, is computed at its midpoint, $\xi_{i}=\xi_{0}+\left(i-1/2\right)\Delta\xi$. The discretized elasticity equation is given by

\begin{equation}
\mathbf{\Pi}_{f}=\mathbf{C\Omega}-\mathbf{S}\label{eq:Discretized_elasticity}
\end{equation}
where $\mathbf{C}$ is the coefficient matrix, and the vector $\mathbf{S}$ accounts for the normal stresses on the fracture (thermally induced stress and far-field stresses). Their values are both given in \citet{LuGo17}.

Taking into account both boundary conditions \eqref{eq:scaled_inlet_bc} and \eqref{eq:scaled_tip_bc}, the fluid flux $\mathit{\check{\Psi}}_{i+1/2}$ at the boundary between the elements $i$ and $i+1$ can be discretized using the Poiseuille law \eqref{eq:scaled_poiseuille}

\begin{equation}
\begin{array}{l}
\mathit{\check{\Psi}}_{1/2}\;\;\;\;\,=\frac{1}{2\pi^{d-1}}\left(1-\frac{\Delta\mathit{\Pi}_{w}}{\Delta\tau}\right),\\
\mathit{\check{\Psi}}_{i+1/2}\;\,=-K_{i}\left(\Pi_{i+1}-\Pi_{i}\right),\;\;\;\;\mathrm{for\ }i=1,\ldots,m-1\\
\mathit{\check{\Psi}}_{m+1/2}=0
\end{array}\label{eq:Discretized_poisuille}
\end{equation}
where $K_{i}=\xi_{i+1/2}^{d-1}\frac{1}{\mathcal{{M}}\Delta\xi}\left(\frac{\mathit{\Omega}_{i}+\mathit{\Omega}_{i+1}}{2}\right)^{3}$ and $\xi_{i+1/2}=\left(\xi_{i}+\xi_{i+1}\right)/{2}$. The continuity equation \eqref{eq:scaled_continuity_eq} is discretized using a backward difference approximation for the time derivatives and a central difference approximation for the spatial derivatives

\begin{equation}
\frac{\mathit{\Omega}_{i}-\mathit{\Omega}_{i}^{0}}{\Delta\tau}+\frac{\mathit{\check{\Psi}}_{i+1/2}-\mathit{\check{\Psi}}_{i-1/2}}{\xi_{i}^{d-1}\Delta\xi}=0\label{eq:Discretized_continuity}
\end{equation}
where $\Omega^{0}$ denotes the known width at the previous time step $\tau^{N-1}$.

\subsection{Coupled nonlinear system}

Combining the the discretized elasticity equation \eqref{eq:Discretized_elasticity}, discretized Poiseuille law \eqref{eq:Discretized_poisuille}, and the discretized continuity equation \eqref{eq:Discretized_continuity}, we derive the final form of the solid-fluid coupled nonlinear system

\begin{equation}
\mathbf{\Xi}\Delta\mathbf{\Omega=\Gamma}\label{eq:Discretized_lubrication}
\end{equation}
in which the matrix $\mathbf{\Xi}$ is given in the component form by $\Xi_{ij}=\delta_{ij}+\frac{\Delta\tau}{\Delta\xi}\frac{1}{\xi_{i}^{d-1}}Z_{ij}$, and

\begin{equation}
\begin{array}{l}
Z_{1j}\;=K_{1}\left(-C_{2j}+C_{1j}\right)+\frac{1}{2\pi^{d-1}\Delta\tau}C_{1j},\;j=1,\ldots,m\\
Z_{ij}\;\,=K_{i}\left(-C_{i+1j}+C_{ij}\right)-K_{i-1}\left(-C_{ij}+C_{i-1j}\right),\;i=2,\ldots,m-1;\;j=1,\ldots,m\\
Z_{mj}=-K_{m-1}\left(-C_{mj}+C_{m-1j}\right),\;j=1,\ldots,m
\end{array}
\end{equation}
and the vector $\mathbf{\Gamma}$ is expressed by

\begin{equation}
\begin{array}{l}
\mathit{\Gamma}_{1}=-\frac{\Delta\tau}{\xi_{1}^{d-1}\Delta\xi}\left[\sum_{j=1}^{m}\left(Z_{1j}-\frac{1}{2\pi^{d-1}\Delta\tau}C_{1j}\right)\mathit{\Omega}_{j}^{0}-K_{1}\left(\varphi_{1}-\varphi_{2}\right)\right]+\ldots \\
+\frac{\Delta\tau}{2\left(\xi_{1}\pi\right)^{d-1}\Delta\xi},\;j=1,\ldots,m\\
\mathit{\Gamma}_{i}=-\frac{\Delta\tau}{\xi_{i}^{d-1}\Delta\xi}\left[\sum_{j=1}^{m}Z_{ij}\mathit{\Omega}_{j}^{0}-K_{i}\left(\varphi_{i}-\varphi_{i+1}\right)-K_{i-1}\left(\varphi_{i}-\varphi_{i-1}\right)\right], \\
i=2,\ldots,m-1;\;j=1,\ldots,m\\
\mathit{\Gamma}_{m}=-\frac{\Delta\tau}{\xi_{m}^{d-1}\Delta\xi}\left[\sum_{j=1}^{m}Z_{mj}\mathit{\Omega}_{j}^{0}-K_{m-1}\left(\varphi_{m}-\varphi_{m-1}\right)\right],\;j=1,\ldots,m
\end{array}
\end{equation}
where $\varphi_{i}$ is given in \citet{LuGo17}.

\subsection{Numerical scheme}

Before the fracture is initiated, we fix the length of the defect ($\gamma_0$) with the initial condition \eqref{eq:scaled_hf_ini_cond}. A constant increment, $\Delta\tau_\text{ini}$, is applied at every step. In this initiation phase, the non-linear system \eqref{eq:Discretized_lubrication} is solved using fixed-point iteration to update the fracture width, $\mathit{\Omega}$, at each time step. Specifically, at iteration $k$, the increment of the opening is found from the solution of the following linear system

\begin{equation}
\mathbf{\Xi}\left(\Delta\mathbf{\Omega}^{k}\right)\Delta\mathbf{\Omega}^{k+1}=\mathbf{\Gamma}\left(\Delta\mathbf{\Omega}^{k}\right)\label{eq:Inner_loop_eq}
\end{equation}

As a result of continuous fluid injection, the fracture width, $\mathit{\Omega}$, keeps increasing until the fracture propagation criterion \eqref{eq:scaled_k_asymptote} is satisfied at the crack tip (i.e., $K_{\mathrm{I}}=K_{\mathrm{IC}}$). Then the algorithm switches to the quasi-static fracture propagation phase, in which a length-controlled algorithm is applied to solve the coupled problem - a fixed length increment is imposed in every step and its corresponding time and width increments are obtained by two nested loops. While the increment of crack width in this phase is still computed by the iterative scheme \eqref{eq:Inner_loop_eq} (inner loop), an additional outer iteration loop based on the global volume balance condition \eqref{eq:scaled_global_vol_bal} is constructed to solve for the unknown time increment $\Delta\tau$ corresponding to the new HF length. Therefore, the outer loop involves checking whether this condition is satisfied, and if not, adjusting $\Delta\tau$ in accordance with global volume balance
\eqref{eq:scaled_global_vol_bal}, which in discretized form is 

\begin{equation}
\Delta\tau^{K}=\Delta\mathit{\Pi}_{w}^{K-1}+2\Delta\xi\sum_{j=1}^{m}\left(\pi\xi_{j}\right)^{d-1}\Delta\mathit{\Omega}_{j}^{K-1}\label{eq:Discretized_glob_vol_bal}
\end{equation}
in which $K$ is the iteration counter for the outer loop.


\end{appendices}


\bibliography{Master_Bib}

\begin{thebibliography}{85}
\providecommand{\natexlab}[1]{#1}
\providecommand{\url}[1]{{#1}}
\providecommand{\urlprefix}{URL }
\providecommand{\doi}[1]{\url{https://doi.org/#1}}
\providecommand{\eprint}[2][]{\url{#2}}
 \bibcommenthead

\bibitem[{Abass et~al(1996)Abass, Hedayati, and Meadows}]{AbHe96}
Abass HH, Hedayati S, Meadows DL (1996) {Nonplanar Fracture Propagation From a
  Horizontal Wellbore: Experimental Study}. SPE Prod Facil 11(03):133--137.
  \doi{10.2118/24823-PA}, \urlprefix\url{https://doi.org/10.2118/24823-PA}

\bibitem[{Abbas and Lecampion(2013)}]{AbLe13}
Abbas S, Lecampion B (2013) Initiation and breakdown of an axisymmetric
  hydraulic fracture transverse to a horizontal wellbore. In: Bunger AP,
  McLennan J, Jeffrey RG (eds) Effective and Sustainable Hydraulic Fracturing.
  Intech, Rijeka, Croatia, chap~19

\bibitem[{Adachi(2001)}]{Adac01}
Adachi J (2001) Fluid-{D}riven {F}racture in {P}ermeable {R}ock. PhD thesis,
  University of Minnesota, Minneapolis, MN

\bibitem[{Batchelor(1967)}]{Batc67}
Batchelor G (1967) An {I}ntroduction to {F}luid {D}ynamics. Cambridge
  University Press, Cambridge UK

\bibitem[{Benouadah et~al(2021)Benouadah, Djabelkhir, Song, Rasouli, and
  Damjanac}]{BeDj21}
Benouadah N, Djabelkhir N, Song X, et~al (2021) Simulation of competition
  between transverse notches versus axial fractures in open hole completion
  hydraulic fracturing. Rock Mech Rock Eng 54:2249--2265.
  \doi{https://doi.org/10.1007/s00603-021-02378-2}

\bibitem[{Brown et~al(2012)Brown, Duchane, Heiken, and Hriscu}]{BrDu12}
Brown DW, Duchane DV, Heiken G, et~al (2012) Mining the earth's heat: hot dry
  rock geothermal energy. Springer Science \& Business Media

\bibitem[{Brudy and Zoback(1999)}]{BrZo99}
Brudy M, Zoback M (1999) Drilling-induced tensile wall-fractures: implications
  for determination of in-situ stress orientation and magnitude. Int J Rock
  Mech Min Sci 36(2):191--215.
  \doi{https://doi.org/10.1016/S0148-9062(98)00182-X},
  \urlprefix\url{https://www.sciencedirect.com/science/article/pii/S014890629800182X}

\bibitem[{Bunger and Detournay(2008)}]{BuDe08}
Bunger AP, Detournay E (2008) Experimental validation of the tip asymptotics
  for a fluid-driven fracture. J Mech Phys Solids 56:3101--3115

\bibitem[{Bunger and Lu(2015)}]{BuLu15}
Bunger AP, Lu G (2015) Time-dependent initiation of multiple hydraulic
  fractures in a formation with varying stresses. Soc Pet Eng J
  20(06):1317--1325

\bibitem[{Bérard and Cornet(2003)}]{BeCo03}
Bérard T, Cornet F (2003) Evidence of thermally induced borehole elongation: a
  case study at soultz, france. Int J Rock Mech Min Sci 40(7):1121--1140.
  \doi{https://doi.org/10.1016/S1365-1609(03)00118-7},
  \urlprefix\url{https://www.sciencedirect.com/science/article/pii/S1365160903001187}

\bibitem[{Cha et~al(2018)Cha, Alqahtani, Yao, Yin, Kneafsey, Wang, Wu, and
  Miskimins}]{ChAl18}
Cha M, Alqahtani NB, Yao B, et~al (2018) Cryogenic fracturing of wellbores
  under true triaxial-confining stresses: experimental investigation. Spe J
  23(04):1271--1289

\bibitem[{Chen and Zhou(2022)}]{ChZh22}
Chen B, Zhou Q (2022) Scaling behavior of thermally driven fractures in deep
  low-permeability formations: A plane strain model with 1-d heat conduction. J
  Geophys Res Solid Earth 127(3):e2021JB022964.
  \doi{https://doi.org/10.1029/2021JB022964},
  \urlprefix\url{https://agupubs.onlinelibrary.wiley.com/doi/abs/10.1029/2021JB022964},
  e2021JB022964 2021JB022964,
  {\href{https://arxiv.org/abs/https://agupubs.onlinelibrary.wiley.com/doi/pdf/10.1029/2021JB022964}{{https://agupubs.onlinelibrary.wiley.com/doi/pdf/10.1029/2021JB022964}}}

\bibitem[{Cheng(2016)}]{Chen16}
Cheng AHD (2016) Poroelasticity. Springer

\bibitem[{Cornet and Valette(1984)}]{CoVa84}
Cornet FH, Valette B (1984) In situ stress determination from hydraulic
  injection test data. J Geophys Res Solid Earth 89(B13):11527--11537.
  \doi{https://doi.org/10.1029/JB089iB13p11527},
  \urlprefix\url{https://agupubs.onlinelibrary.wiley.com/doi/abs/10.1029/JB089iB13p11527},
  {\href{https://arxiv.org/abs/https://agupubs.onlinelibrary.wiley.com/doi/pdf/10.1029/JB089iB13p11527}{{https://agupubs.onlinelibrary.wiley.com/doi/pdf/10.1029/JB089iB13p11527}}}

\bibitem[{Crouch and Starfield(1983)}]{CrSt83}
Crouch S, Starfield A (1983) Boundary {E}lement {M}ethods in {S}olid
  {M}echanics. Unwin Hyman, London

\bibitem[{Daneshy(2011)}]{Dane11}
Daneshy A (2011) {Hydraulic Fracturing of Horizontal Wells: Issues and
  Insights}. In: SPE Hydraulic Fracturing Technology Conference and Exhibition,
  \doi{10.2118/140134-MS}, \urlprefix\url{https://doi.org/10.2118/140134-MS},
  sPE-140134-MS,
  \eprint{https://onepetro.org/SPEHFTC/proceedings-pdf/11HFTC/All-11HFTC/SPE-140134-MS/1689657/spe-140134-ms.pdf}

\bibitem[{{De Bree} and Walters(1989)}]{DeWa89}
{De Bree} P, Walters J (1989) Micro/minifrac test procedures and interpretation
  for in situ stress determination. Int J Rock Mech Min Sci Geomech Abstr
  26(6):515--521. \doi{https://doi.org/10.1016/0148-9062(89)91429-0},
  \urlprefix\url{https://www.sciencedirect.com/science/article/pii/0148906289914290}

\bibitem[{Detournay(2016)}]{Deto16}
Detournay E (2016) Mechanics of hydraulic fractures. Annual Review of Fluid
  Mechanics 48:311--339

\bibitem[{Detournay and Carbonell(1997)}]{DeCa97}
Detournay E, Carbonell R (1997) Fracture-mechanics analysis of the breakdown
  process in minifracture or leakoff test. SPE Production \& Facilities
  August:195--199. {S}PE 28076

\bibitem[{Dobroskok and Ghassemi(2005)}]{DoGh05}
Dobroskok A, Ghassemi A (2005) Crack propagation under thermal influence of a
  wellbore. In: Proceedings of the 30th Workshop on Geothermal Reservoir
  Engineering, Stanford University, CA

\bibitem[{Fernau et~al(2016)Fernau, Lu, Bunger, Prioul, Aidagulov
  et~al}]{FeLu16}
Fernau H, Lu G, Bunger A, et~al (2016) Load-rate dependence of rock tensile
  strength testing: Experimental evidence and implications of kinetic fracture
  theory. In: Proceedings 50th US Rock Mechanics/Geomechanics Symposium, {A}RMA
  16-369

\bibitem[{Ghassemi and Zhang(2004)}]{GhZh04}
Ghassemi A, Zhang Q (2004) Poro-thermoelastic mechanisms in wellbore stability
  and reservoir stimulation. In: Proceedings of the 29th workshop on geothermal
  reservoir engineering, Stanford University, CA

\bibitem[{Ghassemi and Zhang(2006)}]{GhZh06}
Ghassemi A, Zhang Q (2006) Porothermoelastic analysis of the response of a
  stationary crack using the displacement discontinuity method. J Eng Mech
  132(1):26--33. \doi{10.1061/(ASCE)0733-9399(2006)132:1(26)}

\bibitem[{Ghassemi et~al(2007)Ghassemi, Tarasovs, and Cheng}]{GhTa07}
Ghassemi A, Tarasovs S, Cheng AD (2007) A 3-d study of the effects of
  thermomechanical loads on fracture slip in enhanced geothermal reservoirs.
  Int J Rock Mech Min Sci 44(8):1132--1148.
  \doi{https://doi.org/10.1016/j.ijrmms.2007.07.016},
  \urlprefix\url{https://www.sciencedirect.com/science/article/pii/S1365160907001104}

\bibitem[{Ghassemi et~al(2008)Ghassemi, Nygren, and Cheng}]{GhNy08}
Ghassemi A, Nygren A, Cheng A (2008) Effects of heat extraction on fracture
  aperture: A poro–thermoelastic analysis. Geothermics 37(5):525--539.
  \doi{https://doi.org/10.1016/j.geothermics.2008.06.001},
  \urlprefix\url{https://www.sciencedirect.com/science/article/pii/S0375650508000370}

\bibitem[{Griffiths et~al(2018)Griffiths, Lengliné, Heap, Baud, and
  Schmittbuhl}]{GrLe18}
Griffiths L, Lengliné O, Heap MJ, et~al (2018) Thermal cracking in westerly
  granite monitored using direct wave velocity, coda wave interferometry, and
  acoustic emissions. J Geophys Res Solid Earth 123(3):2246--2261.
  \doi{https://doi.org/10.1002/2017JB015191},
  \urlprefix\url{https://agupubs.onlinelibrary.wiley.com/doi/abs/10.1002/2017JB015191},
  {\href{https://arxiv.org/abs/https://agupubs.onlinelibrary.wiley.com/doi/pdf/10.1002/2017JB015191}{{https://agupubs.onlinelibrary.wiley.com/doi/pdf/10.1002/2017JB015191}}}

\bibitem[{Haimson and Fairhurst(1967)}]{HaFa67}
Haimson B, Fairhurst C (1967) Initiation and extension of hydraulic fractures
  in rocks. Soc Pet Eng J pp 310--318. \doi{https://doi.org/10.2118/1710-PA}, ,
  SPE 1710

\bibitem[{Haimson and Cornet(2003)}]{HaCo03}
Haimson BC, Cornet FH (2003) {ISRM} {S}uggested {M}ethods for rock stress
  estimation -- {P}art 3: hydraulic fracturing ({HF}) and/or hydraulic testing
  of pre-existing fracture ({HTPF}). Int J Rock Mech Min Sci 40:1011--1020

\bibitem[{Hubbert and Willis(1957)}]{HuWi57}
Hubbert M, Willis D (1957) Mechanics of hydraulic fracturing. Trans {AIME}
  210:153--168

\bibitem[{Irwin(1957)}]{Irwi57}
Irwin GR (1957) Analysis of stresses and strains near the end of a crack
  transversing a plate. ASME J Appl Mech 24(3):361--364

\bibitem[{Kear and Bunger(2014)}]{KeBu14}
Kear J, Bunger AP (2014) Dependence of static fatigue tests on experimental
  configuration for a crystalline rock. Adv Mater Res 892:863--871.
  {P}roceedings 11th International Fatigue Congress, Melbourne, Australia, 2-7
  March

\bibitem[{Kear et~al(2013)Kear, White, Bunger, Jeffrey, and Hessami}]{KeWh13}
Kear J, White J, Bunger AP, et~al (2013) Three dimensional forms of
  closely-spaced hydraulic fractures. In: Bunger AP, McLennan J, Jeffrey RG
  (eds) Effective and Sustainable Hydraulic Fracturing. Intech, Rijeka,
  Croatia, chap~34

\bibitem[{Keer et~al(1977)Keer, Luk, and Freedman}]{KeLu77}
Keer LM, Luk VK, Freedman JM (1977) Circumferential edge crack in a cylindrical
  cavity. J Appl Mech 44(2):250--254

\bibitem[{Kelkar et~al(2016)Kelkar, WoldeGabriel, and Rehfeldt}]{KeWo15}
Kelkar S, WoldeGabriel G, Rehfeldt K (2016) Lessons learned from the pioneering
  hot dry rock project at fenton hill, usa. Geothermics 63:5--14.
  \doi{https://doi.org/10.1016/j.geothermics.2015.08.008},
  \urlprefix\url{https://www.sciencedirect.com/science/article/pii/S0375650515001091}

\bibitem[{Kirsch(1898)}]{kirs98}
Kirsch G (1898) Die theorie der Elastizit{\"a}t und die Bed{\"u}rfnisse der
  Festigkeitslehre. Springer

\bibitem[{Lakirouhani et~al(2016)Lakirouhani, Detournay, and Bunger}]{LaDe16}
Lakirouhani A, Detournay E, Bunger A (2016) A reassessment of in-situ stress
  determination by hydraulic fracturing. Geophys J Int 205(3):1859--1873

\bibitem[{Lecampion(2012)}]{Leca12}
Lecampion B (2012) Modeling size effects associated with tensile fracture
  initiation from a wellbore. Int J Rock Mech Min Sci 56:67--76

\bibitem[{Lecampion et~al(2013)Lecampion, Abbas, and Prioul}]{LeAb13}
Lecampion B, Abbas S, Prioul R (2013) Competition between transverse and axial
  hydraulic fractures in horizontal wells. In: Proceedings {SPE} Hydraulic
  Fracturing Technology Conference and Exhibition, The Woodlands, Texas, USA,
  {S}PE 163848

\bibitem[{Lecampion et~al(2017)Lecampion, Desroches, Jeffrey, and
  Bunger}]{LeDe17}
Lecampion B, Desroches J, Jeffrey RG, et~al (2017) Experiments versus theory
  for the initiation and propagation of radial hydraulic fractures in
  low-permeability materials. J Geophys Res Solid Earth 122(2):1239--1263

\bibitem[{Lee and Ghassemi(2011)}]{LeGh11}
Lee SH, Ghassemi A (2011) Three-dimensional thermo-poro-mechanical modeling of
  reservoir stimulation and induced microseismicity in geothermal reservoir.
  In: Proceedings of the 36th workshop on geothermal reservoir engineering,
  Stanford University, CA

\bibitem[{Lhomme et~al(2005)Lhomme, Detournay, and Jeffrey}]{LhDe05sfc}
Lhomme T, Detournay E, Jeffrey R (2005) Effect of fluid compressibility and
  borehole radius on the propagation of a fluid-driven fracture. Strength,
  Fracture, and Complexity 3(2-4):149--162

\bibitem[{Li et~al(2020)Li, Ma, Zhang, Zou, Wu, Li, Zhang, and Cao}]{LiMa20}
Li N, Ma X, Zhang S, et~al (2020) Thermal effects on the physical and
  mechanical properties and fracture initiation of laizhou granite during
  hydraulic fracturing. Rock Mech Rock Eng 53:2539--2556

\bibitem[{Li et~al(1998)Li, Cui, and Roegiers}]{LiCu98}
Li X, Cui L, Roegiers JC (1998) Thermoporoelastic analyses of inclined
  boreholes. In: SPE/ISRM Rock Mechanics in Petroleum Engineering, {S}PE--47296

\bibitem[{Liu and Lecampion(2022)}]{LiLe22a}
Liu D, Lecampion B (2022) Laboratory investigation of hydraulic fracture growth
  in zimbabwe gabbro. J Geophys Res Solid Earth 127(11):e2022JB025678.
  \doi{https://doi.org/10.1029/2022JB025678},
  \urlprefix\url{https://agupubs.onlinelibrary.wiley.com/doi/abs/10.1029/2022JB025678}

\bibitem[{Liu et~al(2020)Liu, Lecampion, and Blum}]{LiLe20}
Liu D, Lecampion B, Blum T (2020) {Time-lapse reconstruction of the fracture
  front from diffracted waves arrivals in laboratory hydraulic fracture
  experiments}. Geophys J Int 223(1):180--196. \doi{10.1093/gji/ggaa310},
  \urlprefix\url{https://doi.org/10.1093/gji/ggaa310}

\bibitem[{Lu et~al(2015)Lu, Uwaifo, Ames, Ufondu, Bunger, Prioul, and
  Aidagulov}]{LuUw15}
Lu G, Uwaifo EC, Ames BC, et~al (2015) Experimental demonstration of delayed
  initiation of hydraulic fractures below breakdown pressure in granite. In:
  Proceedings 49th US Rock Mechanics/Geomechanics Symposium, {A}RMA 15-190

\bibitem[{Lu et~al(2017)Lu, Gordeliy, Prioul, and Bunger}]{LuGo17}
Lu G, Gordeliy E, Prioul R, et~al (2017) Modeling initiation and propagation of
  a hydraulic fracture under subcritical conditions. Computer Meth Appl Mech
  Eng 318:61--91

\bibitem[{Lu et~al(2018)Lu, Gordeliy, Prioul, Aidagulov, and Bunger}]{LuGo18}
Lu G, Gordeliy E, Prioul R, et~al (2018) Modeling simultaneous initiation and
  propagation ofmultiple hydraulic fractures under subcritical conditions.
  COMPUT GEOTECH 51(12):3895--3906

\bibitem[{Lu et~al(2020)Lu, Gordeliy, Prioul, Aidagulov, Uwaifo, Ou, and
  Bunger}]{LuGo20}
Lu G, Gordeliy E, Prioul R, et~al (2020) Time-dependent hydraulic fracture
  initiation. J Geophys Res Solid Earth 125(3).
  \doi{https://doi.org/10.1029/2019JB018797}

\bibitem[{Lu et~al(2022{\natexlab{a}})Lu, Momeni, and Lecampion}]{LuMo22}
Lu G, Momeni S, Lecampion B (2022{\natexlab{a}}) Experimental investigation of
  hydraulic fracture growth in an anisotropic rock with pre-existing
  discontinuities under different propagation regimes. In: Proceedings 56th US
  Rock Mechanics/Geomechanics Symposium, {A}RMA 22-239

\bibitem[{Lu et~al(2022{\natexlab{b}})Lu, Zhao, Zheng, and Bunger}]{LuZh22}
Lu G, Zhao W, Zheng J, et~al (2022{\natexlab{b}}) Weakening effect of various
  pore fluids on the tensile strength of granite. Geomech Geophys Geo-energ
  Geo-resour 8(5):144. \doi{https://doi.org/10.1007/s40948-022-00452-9}

\bibitem[{Lu and Cha(2022)}]{LuCh22}
Lu Y, Cha M (2022) Thermally induced fracturing in hot dry rock environments -
  laboratory studies. Geothermics 106:102569.
  \doi{https://doi.org/10.1016/j.geothermics.2022.102569},
  \urlprefix\url{https://www.sciencedirect.com/science/article/pii/S0375650522002152}

\bibitem[{Martin(1993)}]{Mart93}
Martin CD (1993) The strength of massive lac du bonnet granite around
  underground openings. PhD thesis, University of Manitoba, Winnipeg, Manitoba,
  Canada

\bibitem[{McClure and Horne(2014)}]{McHo14}
McClure MW, Horne RN (2014) An investigation of stimulation mechanisms in
  enhanced geothermal systems. Int J Rock Mech Min Sci 72:242--260.
  \doi{https://doi.org/10.1016/j.ijrmms.2014.07.011},
  \urlprefix\url{https://www.sciencedirect.com/science/article/pii/S1365160914002044}

\bibitem[{McTigue(1990)}]{McTi90}
McTigue DF (1990) Flow to a heated borehole in porous, thermoelastic rock:
  Analysis. Water Resour Res 26(8):1763--1774.
  \doi{https://doi.org/10.1029/WR026i008p01763},
  \urlprefix\url{https://agupubs.onlinelibrary.wiley.com/doi/abs/10.1029/WR026i008p01763},
  {\href{https://arxiv.org/abs/https://agupubs.onlinelibrary.wiley.com/doi/pdf/10.1029/WR026i008p01763}{{https://agupubs.onlinelibrary.wiley.com/doi/pdf/10.1029/WR026i008p01763}}}

\bibitem[{Moore et~al(2020)Moore, McLennan, Pankow, Simmons, Podgorney,
  Wannamaker, Jones, Rickard, and Xing}]{MoMc20}
Moore J, McLennan J, Pankow K, et~al (2020) The utah frontier observatory for
  research in geothermal energy (forge): a laboratory for characterizing,
  creating and sustaining enhanced geothermal systems. In: Proceedings of the
  45th Workshop on Geothermal Reservoir Engineering, Stanford University

\bibitem[{Murphy et~al(1981)Murphy, Tester, Grigsby, and Potter}]{MuTe81}
Murphy HD, Tester JW, Grigsby CO, et~al (1981) Energy extraction from fractured
  geothermal reservoirs in low-permeability crystalline rock. J Geophys Res
  Solid Earth 86(B8):7145--7158. \doi{https://doi.org/10.1029/JB086iB08p07145},
  \urlprefix\url{https://agupubs.onlinelibrary.wiley.com/doi/abs/10.1029/JB086iB08p07145},
  {\href{https://arxiv.org/abs/https://agupubs.onlinelibrary.wiley.com/doi/pdf/10.1029/JB086iB08p07145}{{https://agupubs.onlinelibrary.wiley.com/doi/pdf/10.1029/JB086iB08p07145}}}

\bibitem[{Nasseri et~al(2009)Nasseri, Tatone, Grasselli, and Young}]{NaTa09}
Nasseri M, Tatone B, Grasselli G, et~al (2009) Fracture toughness and fracture
  roughness interrelationship in thermally treated westerly granite. Pure Appl
  Geophys 166:801--822. \doi{https://doi.org/10.1007/s00024-009-0476-3}

\bibitem[{Nilson and Proffer(1984)}]{NiPr84}
Nilson R, Proffer WJ (1984) Engineering formulas for fractures emanating from
  cylindrical and spherical holes. ASME J Appl Mech 51:929--933

\bibitem[{Nowacki(1986)}]{Nowa86}
Nowacki W (1986) Thermoelasticity, 2nd edn. Pergamon Press

\bibitem[{Perkins and Gonzalez(1984)}]{PeGo84}
Perkins TK, Gonzalez JA (1984) {Changes in Earth Stresses Around a Wellbore
  Caused by Radially Symmetrical Pressure and Temperature Gradients}. SPE J
  24(02):129--140. \doi{10.2118/10080-PA},
  \urlprefix\url{https://doi.org/10.2118/10080-PA}

\bibitem[{Rice(1972)}]{rice72}
Rice J (1972) Some remarks on elastic crack-tip stress fields. Int J Solids
  Struct 8(6):751--758

\bibitem[{Savitski and Detournay(2002)}]{SaDe02}
Savitski A, Detournay E (2002) Propagation of a penny-shaped fluid-driven
  fracture in an impermeable rock: asymptotic solutions. Int J Solids Struct
  39:6311--6337

\bibitem[{Sinha and Joshi(2011)}]{SiJo11}
Sinha A, Joshi YK (2011) {Downhole Electronics Cooling Using a Thermoelectric
  Device and Heat Exchanger Arrangement}. J Electron Packag 133(4).
  \doi{10.1115/1.4005290}, \urlprefix\url{https://doi.org/10.1115/1.4005290},
  041005

\bibitem[{Stehfest(1970)}]{Steh70}
Stehfest H (1970) Numerical inversion of {L}aplace transforms. Commun ACM
  13:47--49 and 624

\bibitem[{Stephens and Voight(1982)}]{StVo82}
Stephens G, Voight B (1982) Hydraulic fracturing theory for conditions of
  thermal stress. Int J Rock Mech Min Sci Geomech Abstr 19(6):279--284.
  \doi{https://doi.org/10.1016/0148-9062(82)91364-X},
  \urlprefix\url{https://www.sciencedirect.com/science/article/pii/014890628291364X}

\bibitem[{Tao and Ghassemi(2010)}]{TaGh10}
Tao Q, Ghassemi A (2010) Poro-thermoelastic borehole stress analysis for
  determination of the in situ stress and rock strength. Geothermics
  39(3):250--259. \doi{https://doi.org/10.1016/j.geothermics.2010.06.004},
  \urlprefix\url{https://www.sciencedirect.com/science/article/pii/S0375650510000258}

\bibitem[{Tarasovs and Ghassemi(2011)}]{TaGh11}
Tarasovs S, Ghassemi A (2011) Propagation of a system of cracks under thermal
  stress. In: Proceedings 45th US Rock Mechanics/Geomechanics Symposium, {A}RMA
  11-58

\bibitem[{Tarasovs and Ghassemi(2012{\natexlab{a}})}]{TaGh12b}
Tarasovs S, Ghassemi A (2012{\natexlab{a}}) On the role of thermal stress in
  reservoir stimulation. In: Proceedings of the 37th Workshop on Geothermal
  Reservoir Engineering, Stanford University, CA

\bibitem[{Tarasovs and Ghassemi(2012{\natexlab{b}})}]{TaGh12a}
Tarasovs S, Ghassemi A (2012{\natexlab{b}}) Radial cracking of a borehole by
  pressure and thermal shock. In: Proceedings 46th US Rock
  Mechanics/Geomechanics Symposium, {A}RMA 12-425

\bibitem[{Tarasovs and Ghassemi(2014)}]{TaGh14}
Tarasovs S, Ghassemi A (2014) Self-similarity and scaling of thermal shock
  fractures. Phys Rev E 90:012403. \doi{10.1103/PhysRevE.90.012403},
  \urlprefix\url{https://link.aps.org/doi/10.1103/PhysRevE.90.012403}

\bibitem[{Tomac and Gutierrez(2017)}]{ToGu17}
Tomac I, Gutierrez M (2017) Coupled hydro-thermo-mechanical modeling of
  hydraulic fracturing in quasi-brittle rocks using bpm-dem. J Rock Mech
  Geotech Eng 9(1):92--104. \doi{https://doi.org/10.1016/j.jrmge.2016.10.001},
  \urlprefix\url{https://www.sciencedirect.com/science/article/pii/S1674775516302116}

\bibitem[{Tran et~al(2010)Tran, Roegiers, and Thiercelin}]{TrRo10}
Tran DT, Roegiers JC, Thiercelin M (2010) Thermally-induced tensile fractures
  in the barnett shale and their implications to gas shale fracability. In:
  Proceedings 44th US Rock Mechanics/Geomechanics Symposium, {A}RMA 10-466

\bibitem[{Wang et~al(2022)Wang, Bai, Wang, Wei, and Liang}]{WaBa22}
Wang D, Bai B, Wang B, et~al (2022) Impacts of fracture roughness and
  near-wellbore tortuosity on proppant transport within hydraulic fractures.
  Sustainability 14(14). \doi{10.3390/su14148589},
  \urlprefix\url{https://www.mdpi.com/2071-1050/14/14/8589}

\bibitem[{Wang et~al(2023)Wang, Tang, Du, Zhang, Hou, and Tang}]{WaTa23}
Wang X, Tang M, Du X, et~al (2023) Three-dimensional experimental and numerical
  investigations on fracture initiation and propagation for oriented
  limited-entry perforation and helical perforation. Rock Mech Rock Eng
  56:437--462

\bibitem[{Wang and Papamichos(1994)}]{WaPa94}
Wang Y, Papamichos E (1994) Conductive heat flow and thermally induced fluid
  flow around a well bore in a poroelastic medium. Water Resour Res
  30(12):3375--3384. \doi{https://doi.org/10.1029/94WR01774},
  \urlprefix\url{https://agupubs.onlinelibrary.wiley.com/doi/abs/10.1029/94WR01774},
  {\href{https://arxiv.org/abs/https://agupubs.onlinelibrary.wiley.com/doi/pdf/10.1029/94WR01774}{{https://agupubs.onlinelibrary.wiley.com/doi/pdf/10.1029/94WR01774}}}

\bibitem[{Waters et~al(2006)Waters, Heinze, Jackson, Ketter, Daniels, and
  Bentley}]{WaHe06}
Waters G, Heinze J, Jackson R, et~al (2006) {Use of Horizontal Well Image Tools
  To Optimize Barnett Shale Reservoir Exploitation}. In: SPE Annual Technical
  Conference and Exhibition, \doi{10.2118/103202-MS},
  \urlprefix\url{https://doi.org/10.2118/103202-MS}, sPE-103202-MS

\bibitem[{Weijers(1995)}]{Weij95}
Weijers L (1995) The near-wellbore geometry of hydraulic fractures initiated
  from horizontal and deviated wells. PhD thesis, TU Delft, Delft, Netherlands

\bibitem[{Weijers et~al(1994)Weijers, de~Pater, Owens, and Kogsb?ll}]{WeDe94}
Weijers L, de~Pater CJ, Owens KA, et~al (1994) {Geometry of Hydraulic Fractures
  Induced From Horizontal Wellbores}. SPE Prod Facil 9(02):87--92.
  \doi{10.2118/25049-PA}, \urlprefix\url{https://doi.org/10.2118/25049-PA},
  {\href{https://arxiv.org/abs/https://onepetro.org/PO/article-pdf/9/02/87/2606031/spe-25049-pa.pdf}{{https://onepetro.org/PO/article-pdf/9/02/87/2606031/spe-25049-pa.pdf}}}

\bibitem[{Winner et~al(2018)Winner, Lu, Prioul, Aidagulov, and Bunger}]{WiLu18}
Winner RA, Lu G, Prioul R, et~al (2018) Acoustic emission and kinetic fracture
  theory for time-dependent breakage of granite. Eng Fracture Mech
  199:101--113. \doi{https://doi.org/10.1016/j.engfracmech.2018.05.004}

\bibitem[{Xiao et~al(2022)Xiao, Hu, Meng, Li, Wang, and Chen}]{XiHu22}
Xiao D, Hu Y, Meng Y, et~al (2022) Research on wellbore temperature control and
  heat extraction methods while drilling in high-temperature wells. J Pet Sci
  Eng 209:109814. \doi{https://doi.org/10.1016/j.petrol.2021.109814},
  \urlprefix\url{https://www.sciencedirect.com/science/article/pii/S0920410521014339}

\bibitem[{Xing et~al(2022)Xing, Damjanac, Moore, and McLennan}]{XiDa21}
Xing P, Damjanac B, Moore J, et~al (2022) Flowback test analyses at the utah
  frontier observatory for research in geothermal energy (forge) site. Rock
  Mech Rock Eng 55:3023--3040. \doi{https://doi.org/10.1007/s00603-021-02604-x}

\bibitem[{Zhang et~al(2011)Zhang, Jeffrey, Bunger, and Thiercelin}]{ZhJe11}
Zhang X, Jeffrey RG, Bunger AP, et~al (2011) Initiation and growth of a
  hydraulic fracture from a circular wellbore. Int J Rock Mech Min Sci
  48:984--995

\bibitem[{Zhou et~al(2021)Zhou, Jin, Zhuang, Xin, and Zhang}]{ZhJi21}
Zhou Z, Jin Y, Zhuang L, et~al (2021) Pumping rate-dependent temperature
  difference effect on hydraulic fracturing of the breakdown pressure in hot
  dry rock geothermal formations. Geothermics 96:102175

\bibitem[{Zimmermann et~al(2010)Zimmermann, Moeck, and Blöcher}]{ZiMo10}
Zimmermann G, Moeck I, Blöcher G (2010) Cyclic waterfrac stimulation to
  develop an enhanced geothermal system {(EGS)} - conceptual design and
  experimental results. Geothermics 39(1):59--69.
  \doi{https://doi.org/10.1016/j.geothermics.2009.10.003},
  \urlprefix\url{https://www.sciencedirect.com/science/article/pii/S0375650509000674}

\end{thebibliography}

\end{document}